\documentclass[usegraphicx,useAMS,usenatbib]{mn2e}
 
\usepackage{times}
\usepackage[fleqn]{amsmath}
\usepackage{amssymb}

\newcommand{\plotone}[1]{%
  \includegraphics[width=\columnwidth]{#1}}
\newcommand{\plottwo}[1]{%
  \includegraphics[width=\textwidth]{#1}}

\def\kms{km\,s$^{-1}$}
\def\dgr{$^\circ$}

\def\Msun{M$_\odot$}

\def\Lsunpcsq{L$_{\odot,I}$\,pc$^{-2}$}
\def\MLsun{M$_\odot$/L$_\odot$}

\def\farcs{\hbox{$.\!\!^{\prime\prime}$}}
\def\arcs{\hbox{$^{\prime\prime}$}}
\def\Sauron{\texttt{SAURON}}
\usepackage{journalnames}

\title[Triaxial orbit based models]{Triaxial orbit based galaxy models 
with an application to the (apparent) decoupled core galaxy NGC~4365} 
\author[van den Bosch et al.]{R. C. E. van den
Bosch$^1$\thanks{E-mail: bosch@strw.leidenuniv.nl}, G. van de Ven$^{1,2,3}$\thanks{Hubble Fellow},
E. K. Verolme$^{1,4}$, 
\newauthor
M. Cappellari$^{1,5}$, P. T. de Zeeuw$^{1,6}$ \\ 
$^1$ Sterrewacht Leiden, Universiteit Leiden, Postbus 9513, 2300 RA Leiden, The Netherlands \\
$^2$Department of Astrophysical Sciences, Peyton Hall, Princeton, NJ
08544, USA\\
$^3$ Institute for Advanced Study, Einstein Drive, NJ 08540, Princeton, USA\\
$^4$ TNO Defense, Security and Safety, Lange Kleiweg 137,
2280 AA, Rijswijk, The Netherlands\\
$^5$ Sub-Department of Astrophysics, University of Oxford, Denys Wilkinson Building, Keble Road, Oxford, OX1 3RH, England\\
$^6$ European Southern Observatory, D-85748 Garching bei M\"unchen\\
}

\date{Draft $ $Date: 2007/12/03 13:52:17 $ $Revision: 1.858 $ $}

\pagerange{\pageref{firstpage}--\pageref{lastpage}} \pubyear{2007}
\begin{document}
\maketitle
\label{firstpage}

\begin{abstract}
  We present a flexible and efficient method to construct triaxial
  dynamical models of galaxies with a central black hole, using
  Schwarzschild's orbital superposition approach. Our method is
  general and can deal with realistic luminosity distributions, which
  project to surface brightness distributions that may show position
  angle twists and ellipticity variations. The models are fit to
  measurements of the full line-of-sight velocity distribution
  (wherever available). We verify that our method is able to reproduce
  theoretical predictions of a three-integral triaxial Abel model.  In
  a companion paper (\nocite{2006vdV06}{van de Ven}, {de Zeeuw} \& {van den  Bosch}), we demonstrate that the
  method recovers the phase-space distribution function. We
  apply our method to two-dimensional observations of the E3 galaxy
  NGC~4365, obtained with the integral-field spectrograph \Sauron, and
  study its internal structure, showing that the observed
  kinematically decoupled core is not physically distinct from the
  main body and the inner region is close to oblate axisymmetric.
\end{abstract}

\begin{keywords} galaxies: elliptical and lenticular, cD - galaxies:
kinematics and dynamics - galaxies: structure
\end{keywords}

\section{Introduction}

\nocite{1976MNRAS.177...19B,1978MNRAS.183..501B}{Binney} (1976, 1978) argued convincingly
that elliptical galaxies may well have triaxial intrinsic shapes,
based on the observed slow rotation of the stars
\nocite{1975ApJ...200..439B,1977ApJ...218L..43I}({Bertola} \& {Capaccioli} 1975; {Illingworth} 1977), the presence of
isophote twists in the surface brightness distribution
\nocite{1979ApJ...227...56W}(e.g., {Williams} \& {Schwarzschild} 1979), the presence of velocity
gradients along the apparent minor axis \nocite{1978AJ.....83.1360S}(`minor-axis
rotation', {Schechter} \& {Gunn} 1978), and evidence from N-body
simulations \nocite{1978MNRAS.185..227A}({Aarseth} \& {Binney} 1978). Schwarzschild's
\nocite{1979ApJ...232..236S, 1982ApJ...263..599S} (1979, 1982) numerical
models demonstrated that such systems can be in dynamical equilibrium,
and suggested that their observed kinematics can be rich \nocite{1991AJ....102..882S}(see
also, e.g., {Statler} 1991). This is supported by the
discovery of kinematically decoupled cores in the late
nineteen-eighties \nocite{1988A&A...202L...5B, 1988ApJ...327L..55F}({Bender} 1988; {Franx} \& {Illingworth} 1988)
and, more recently, by observations with integral-field spectrographs
such as \Sauron, which reveal that some ellipticals have
point-symmetric rather than bi-symmetric velocity fields, and often
contain kinematically decoupled components
\nocite{2004MNRAS.352..721E}(e.g., {Emsellem} {et~al.} 2004). This means that these galaxies
are not axisymmetric.\looseness=-2

Subsequent work on triaxial dynamical models focused mostly on models
with a cusp in the central density profile, on the effect of a central
black hole, and on the range of shapes for which triaxial models could
be in dynamical equilibrium \nocite{1985MNRAS.216..467G,
1987ApJ...314..476L,1987ApJ...321..113S, 1992ApJ...389...79H,
1993ApJ...409..563S,1996ApJ...460..136M, 1998PhDT........14S,
2002PhDT.........8T}(e.g., {Gerhard} \& {Binney} 1985; {Levison} \& {Richstone} 1987; {Statler} 1987; {Hunter} \& {de Zeeuw} 1992; {Schwarzschild} 1993; {Merritt} \& {Fridman} 1996; {Siopis} 1998; {Terzi{\'c}} 2002). With the exception of studies of the Galactic
Bulge \nocite{1996MNRAS.283..149Z, 2000MNRAS.314..433H}({Zhao} 1996; {H{\"a}fner} {et~al.} 2000), most of this
work was restricted to finding (numerical) distribution functions
consistent with a given triaxial density. This showed that many
different distribution functions may reproduce the same triaxial
density, and that these dynamical models all have different observable
kinematic properties, but detailed comparison to observations received
little attention \nocite{1994MNRAS.271..924A,
1999MNRAS.303..455M}({Arnold}, {de Zeeuw} \&  {Hunter} 1994; {Mathieu} \& {Dejonghe} 1999). Ad-hoc kinematic models were used to constrain
the distribution of intrinsic shapes \nocite{1985MNRAS.212..767B,
1991ApJ...383..112F}({Binney} 1985; {Franx}, {Illingworth} \& {de  Zeeuw} 1991) or of individual objects
\nocite{1993A&A...275...61T, 1994ApJ...425..500S, 
2001AJ....121..244S, 1999AJ....117..126S, 2004MNRAS.353....1S}({Tenjes} {et~al.} 1993; {Statler} 1994b, 2001; {Statler}, {Dejonghe} \&  {Smecker-Hane} 1999; {Statler} {et~al.} 2004).

The possibility to measure accurate line-of-sight velocity
distributions (LOSVDs) in elliptical galaxies from observations of the
stellar absorption lines \nocite{1990A&A...229..441B,
1993ApJ...407..525V}(e.g., {Bender} 1990; {van der Marel} \& {Franx} 1993), and the realization that these are the only way
to distinguish radial variations in mass-to-light-ratio $M/L$ from
radial variations in the anisotropy of the orbital structure
\nocite{1987MNRAS.224...13D, 1993MNRAS.265..213G}({Dejonghe} 1987; {Gerhard} 1993), led to the
development of detailed spherical, and subsequently axisymmetric,
numerical dynamical models aimed to fit all these kinematic
measurements. These are generally constructed with a variant of
Schwarzschild's \nocite{1979ApJ...232..236S} (1979) orbit superposition
method, in which occupation numbers are found for a representative
library of orbits calculated in the gravitational potential of the
galaxy. The aim is to measure the mass of the central black holes
\nocite{1998ApJ...493..613V, 2003ApJ...583...92G, 2004ApJ...602...66V,
2005ApJ...628..137V, 2006ApJ...641..852V, 2006MNras...shapiro}({van der Marel} {et~al.} 1998; {Gebhardt} {et~al.} 2003; {Valluri}, {Merritt} \&  {Emsellem} 2004; {Valluri} {et~al.} 2005; {van den Bosch} {et~al.} 2006; {Shapiro} {et~al.} 2006), to
deduce the properties of dark halos
\nocite{1997ApJ...488..702R, 1999ApJS..124..383C,
2001AJ....121.1936G, 2005MNRAS.360.1355T}(e.g., {Rix} {et~al.} 1997; {Cretton} {et~al.} 1999; {Gerhard} {et~al.} 2001; {Thomas} {et~al.} 2005), and to derive the internal
orbital structure and intrinsic shape \nocite{2002MNRAS.335..517V,
2002ApJ...578..787C, 2006MNRAS.366.1126C, 2005MNRAS.357.1113K,
2006A&A...445..513V}({Verolme} {et~al.} 2002; {Cappellari} {et~al.} 2002, 2006; {Krajnovi{\'c}} {et~al.} 2005; {van de Ven} {et~al.} 2006). Some galaxies display significant signatures of
non-axisymmetry, suggesting they are intrinsically triaxial. 

The logical next step is to construct realistic triaxial models, which
fit the details of the observed surface brightness, including isophote
twists, nuclear stellar disks and a central cusp, as well as the
two-dimensional kinematic measurements. This is a non-trivial
undertaking, as the parameter range to be explored for a given model
is significantly larger than in axisymmetric geometry, and the
internal dynamical structure is more complicated, as it includes four
major orbit families, a host of minor families and chaotic orbits.
However, the ability to construct such models will make it possible to
derive reliable intrinsic parameters for giant elliptical galaxies,
and opens the way for a systematic exploration of their properties. In
this paper, we describe a practical method for doing this, and report
an application which accurately reproduces the two-dimensional
kinematic measurements of the triaxial E3 galaxy NGC~4365, obtained
with \Sauron.  In the companion paper \nocite{2006vdV06}({van de Ven} {et~al.} 2007, hereafter
vdV07) we apply the method to analytical triaxial
three-integral models and show that it reliably recovers the input
three-integral distribution function.

We start with a short section on Schwarzschild's method
(\S\ref{schwarzschild}), which includes a brief overview of our
implementation. We then give a step-by-step description of the main
properties of our formalism
(\S\ref{MGEsection}--\ref{superposition}). In \S\ref{sec:test} we test
the method, including the ability to recover the global input
parameters. We construct a triaxial model for NGC~4365 in
\S\ref{ngc4365}, and we summarize our conclusions in
\S\ref{discussion}.

\section{Schwarzschild's method}
\label{schwarzschild}
\subsection{Brief historical overview}
Schwarzschild's \nocite{1979ApJ...232..236S} (1979) orbit superposition
method is a flexible method to build dynamical models of early-type
galaxies. The original implementation was aimed at reproducing a given
triaxial density distribution. Subsequently, it was applied to a large
variety of density distributions, from spherically and axially
symmetric \nocite{1980ApJ...238..103R, 1982ApJ...252..496R,
1984ApJ...281..100R, 1984ApJ...286...27R, 1985ApJ...295..340L,
2004ApJ...602...66V}({Richstone} 1980, 1982, 1984; {Richstone} \& {Tremaine} 1984; {Levison} \& {Richstone} 1985; {Valluri} {et~al.} 2004) to triaxial shapes
\nocite{1982ApJ...263..599S, 1993ApJ...409..563S,
1986ApJ...306...48V, 1987ApJ...314..476L, 1987ApJ...321..113S,
1996ApJ...460..136M, 2000MNRAS.319...43S, 2000MNRAS.314..433H}(e.g., {Schwarzschild} 1982, 1993; {Vietri} 1986; {Levison} \& {Richstone} 1987; {Statler} 1987; {Merritt} \& {Fridman} 1996; {Siopis} \& {Kandrup} 2000; {H{\"a}fner} {et~al.} 2000).

\nocite{1984A&A...141..171P}{Pfenniger} (1984) showed that it is possible to include
measurements of the mean line-of-sight velocity and the second
velocity moment, provided that the true second velocity moment
$\langle v^2\rangle$ is used and not the velocity dispersion
$\sigma^2=\langle v^2\rangle -\langle v \rangle ^2$. The reason for
this requirement is that the dispersion depends quadratically on the
first velocity moment and can therefore not be included in a linear
orbit superposition method \nocite{1989ApJ...343..113D}(but see {Dejonghe} 1989).
\nocite{1996MNRAS.283..149Z}{Zhao} (1996) used this principle to build triaxial
models of the Galactic Bulge. At the same time, theoretical
\nocite{1987MNRAS.224...13D}({Dejonghe} 1987) and observational
\nocite{1988ApJ...327L..55F}({Franx} \& {Illingworth} 1988) investigations showed that LOSVDs are
generally not Gaussian-shaped and higher-order velocity moments are
required to describe the true profile. This stimulated the use of the
so-called Gauss-Hermite (GH) moments \nocite{1993ApJ...407..525V,
  1993MNRAS.265..213G}({van der Marel} \& {Franx} 1993; {Gerhard} 1993).

The first implementations of Schwarzschild's method that used
additional kinematic information were designed for the modeling of
spherical galaxies \nocite{1984ApJ...286...27R, 1997ApJ...488..702R}({Richstone} \& {Tremaine} 1984; {Rix} {et~al.} 1997).
Orbits in these models obey four integrals of motion: the energy $E$
and all three components of the angular momentum
$\mathbf{L}=(L_x,L_y,L_z)$. While useful, this software was still of
limited applicability, as most galaxies are not round, but
axisymmetric or triaxial. Orbits in oblate axisymmetric galaxies
conserve at least the two classical integrals $E$ and $L_z$ (which is
the component of the angular momentum along the short axis), while it
has been known for a long time that most orbits in our Galaxy conserve
an additional non-classical third integral of motion
\nocite{1960ZA.....49..273C, 1962BAN....16..241O}(e.g., {Contopoulos} 1960; {Ollongren} 1962). A more
general version of the Schwarzschild software was therefore developed
to model axisymmetric galaxies with three-integral distribution
functions \nocite{1998ApJ...493..613V, 1999ApJS..124..383C,
  2005MNRAS.360.1355T}({van der Marel} {et~al.} 1998; {Cretton} {et~al.} 1999; {Thomas} {et~al.} 2005). Results that were obtained with the extended
Schwarzschild method indeed showed that the third integral is an
essential ingredient of realistic axisymmetric galaxy models
\nocite{1998ApJ...493..613V, 2002MNRAS.335..517V}({van der Marel} {et~al.} 1998; {Verolme} {et~al.} 2002), that we can derive
information on the phase-space structure of galaxies
\nocite{2002ApJ...578..787C,2005MNRAS.357.1113K}({Cappellari} {et~al.} 2002; {Krajnovi{\'c}} {et~al.} 2005), that we can use the
method to measure the mass of the central black hole in galaxies
\nocite{2003ApJ...583...92G}({Gebhardt} {et~al.} 2003) and that proper motion kinematic
observations can be used \nocite{2006A&A...445..513V}({van de Ven} {et~al.} 2006), provided that
the models have sufficient internal freedom, e.g., the total number of
orbits is large enough \nocite{2004ApJ...602...66V,
  2004MNRAS.353..391T, 2004MNRAS.347L..31C, 2004astro.ph..3257R,
  2006MNRAS.373..425M}({Valluri} {et~al.} 2004; {Thomas} {et~al.} 2004; {Cretton} \& {Emsellem} 2004; {Richstone} {et~al.} 2004; {Magorrian} 2006).

\subsection{Generalization to triaxial geometry}
\label{implementation}

The method described here uses many of the ideas and algorithms
described in \nocite{1997ApJ...488..702R}{Rix} {et~al.} (1997), \nocite{1998ApJ...493..613V}{van der Marel} {et~al.} (1998),
\nocite{ 1999ApJS..124..383C} (), \nocite{ 2002MNRAS.335..517V} () and \nocite{
  2006MNRAS.366.1126C} (). The computer program for triaxial geometry was
written largely from scratch.

The standard implementation of the extended Schwarzschild method
starts from a surface brightness distribution, which we parametrise
with a sum of Gaussians (\S\ref{mge}). The intrinsic mass distribution
and potential are then obtained by deprojecting the surface density,
which requires a choice for the viewing angle(s) along which the
object is observed (\S\ref{mgedeproj}). The potential calculation is
outlined in \S\ref{potacc}.

In the potential, the initial conditions for a representative orbit
library are found (\S\ref{orbits}). These orbital components must
include all types of orbits that the potential supports, to avoid any
bias \nocite{2004MNRAS.353..391T}(e.g., {Thomas} {et~al.} 2004).

Schwarzschild's method tries to find a steady-state model of a galaxy,
requiring orbital building blocks to be time-independent. We integrate
the orbits for a fixed time of 200 times the period of a closed
elliptical orbit with the same energy.

During orbit integration, the intrinsic and projected properties are
stored on grids, in order to allow for comparison with the data
(\S\ref{storage grids}). The quantities that will be compared to
observations are spatially convolved with the same point spread
function (PSF) as the observations.

After orbit integration, the superposition of orbits whose properties
best match the observational data is determined.  The superposition
can be constructed by using linear or quadratic programming
\nocite{1979ApJ...232..236S, 1982ApJ...263..599S, 1993ApJ...409..563S,
  1984ApJ...287..475V, 1989ApJ...343..113D}({Schwarzschild} 1979, 1982, 1993; {Vandervoort} 1984; {Dejonghe} 1989), maximum entropy methods
\nocite{1988ApJ...327...82R, 2003ApJ...583...92G, 2004MNRAS.353..391T}({Richstone} \& {Tremaine} 1988; {Gebhardt} {et~al.} 2003; {Thomas} {et~al.} 2004)
or with a least-squares solver as was used in many of the axisymmetric
three-integral implementations \nocite{1997ApJ...488..702R,
  1998ApJ...493..613V, 2006MNRAS.366.1126C}({Rix} {et~al.} 1997; {van der Marel} {et~al.} 1998; {Cappellari} {et~al.} 2006). Here we use the a
quadratic programming solver as it finds the best-fitting
superposition in a least squares sense, while allowing for additional
contraints (\S\ref{qp}).

\section{Mass parameterization, potential and accelerations}
\label{MGEsection}
In this section, we describe the method that we use to obtain a
triaxial mass model from the observed surface brightness. We describe
a convenient mass parametrisation and derive the corresponding
potential and accelerations. A summary of symbols introduced in this
section is given in Table~\ref{tab:symbolsummary}.

\subsection{The MGE parameterization}
\label{mge}
In order to derive the intrinsic luminosity density from the observed
galaxy surface brightness, a deprojection is required. For a spherical
galaxy, this leads to a unique solution \nocite{1987gady.book.....B}({Binney} \& {Tremaine} 1987).
This is not the case for an axisymmetric object, unless it is seen
edge-on \nocite{1987IAUS..127..397R}({Rybicki} 1987). This non-uniqueness is even
stronger for triaxial shapes, where the deprojection is not unique
from any viewing direction \nocite{1996sgni.conf..138G}(e.g., {Gerhard} 1996). For
this reason, the assumption that an object is triaxial is not
sufficient to uniquely recover the intrinsic luminosity density from
an observed image, and additional assumptions have to be made.

The simplest option is to assume that the intrinsic density is
stratified on similar triaxial ellipsoids. The isophotes that are
produced by such a mass model are similar coaxial ellipses
\nocite{1956ApJ...124..643C, 1977ApJ...213..368S}({Contopoulos} 1956; {Stark} 1977), which is
approximately consistent with observations of some galaxies. However,
many objects display position angle twists and ellipticity variations,
which cannot be reproduced by these simple models. More flexible mass
models are therefore required to reproduce these observed features.

A general approach to the triaxial deprojection problem would be to
use fully non-parametric methods
\nocite{1992mde..book.....S}(e.g., {Scott} 1992). This has already been done in
the axisymmetric case by \nocite{1997MNRAS.287...35R}{Romanowsky} \& {Kochanek} (1997) and in the
triaxial case by \nocite{2002MNRAS.330..591B}{Bissantz} \& {Gerhard} (2002). Unfortunately, these
methods are complicated, require a significant amount of time before
convergence is reached, and do not always provide a global solution.

We therefore decided to parameterize the mass distribution by using a
Multi-Gaussian Expansion \nocite{1992A&A...253..366M,
  1994A&A...285..723E, 2002MNRAS.333..400C}(MGE; {Monnet}, {Bacon} \&  {Emsellem} 1992; {Emsellem}, {Monnet}, \&  {Bacon} 1994; {Cappellari} 2002). We assume that the
intrinsic density can be described as a sum of coaxial triaxial
Gaussian distributions. The Gaussians do not constitute a complete
basis of functions and therefore cannot reproduce any arbitrary
positive density distribution. However, MGE models can reproduce a
large variety of densities, which appears realistic when projected
along any viewing direction, including mass models with radially
varying triaxiality, multiple photometric components and disks.

Accordingly, we write the triaxial MGE luminosity density as
\begin{multline}
  \label{density}
  \rho(x,y,z) = \sum_{j=1}^N \;(M/L)\;
  \frac{L_j}{(\sigma_j\sqrt{2\pi})^3 p_j q_j}
  \\ \times \exp\left[
    -\frac{1}{2\sigma_j^2}
    \left(x^2\!+\!\frac{y^2}{p_j^2} + \frac{z^2}{q_j^2} \right)
  \right],
\end{multline}
where $N$ is the number of required Gaussian components, $L_j$ is the
luminosity of the $j$th Gaussian, $p_j$ and $q_j$ are the axial
ratios, and $\sigma_j$ is the corresponding dispersion along the
$x$-axis. Moreover, $M/L$ is the mass-to-light ratio, and $(x,y,z)$ is
a system of coordinates centered on the common origin of the Gaussians
and aligned with the common principal axes of the Gaussians.

\subsection{Transformation from intrinsic to projected coordinates}
To be able to compute the projection of the density in
eq.~\eqref{density} on the sky-plane, we introduce a new coordinate
system, $(x',y',z')$ as defined in \nocite{1985MNRAS.212..767B}{Binney} (1985). Here,
$z'$ is located along the line-of-sight and $x'$ is in the
$(x,y)$-plane.

To go between these coordinate systems two transformatons are needed.
First, a projection to the sky-plane given by a projection matrix
\begin{equation}
  \label{eqprojtointr}
  \mathbf{P} =
  \begin{pmatrix}
    -\sin\varphi           & \cos\varphi            & 0 \\
    -\cos\vartheta\cos\varphi & -\cos\vartheta\sin\varphi & \sin\vartheta \\
     \sin\vartheta\cos\varphi & \sin\vartheta\sin\varphi  & \cos\vartheta
  \end{pmatrix},
\end{equation}
where the two usual spherical coordinates $(\vartheta,\varphi)$ define
the orientation of the line-of-sight with respect to the principal
axes of the object. For example, (90\dgr,0\dgr), (90\dgr,90\dgr),
(0\dgr,0\dgr...90\dgr) are the views down the long, intermediate and
short axis, respectively. Secondly, a rotation on the sky-plane is
given by the matrix
\begin{equation}
  \label{eqprojpsi}
  \mathbf{R} =
  \begin{pmatrix}
     \sin\psi & -\cos\psi & 0 \\
     \cos\psi &  \sin\psi & 0 \\
     0        &  0        & 1
  \end{pmatrix}.
\end{equation}
The angle $\psi$ is required to specify the rotation of the object
around the line-of-sight. The rotation $\psi$ is chosen to align the
major axis of the projected ellipse (of the innermost MGE component,
see eq.~\eqref{eq:defpsi} below) with the $x'$-axis. For an oblate
axisymmetric intrinsic shape $\psi$ equals $90$\dgr.

\subsection{The observed surface brightness of an MGE}
\label{mgedeproj}

The projected surface brightness (SB) that corresponds to the density
of eq.~\eqref{density} can be written as a sum of two-dimensional
Gaussians of the form:
\begin{equation}
  \label{surf_twist}
  \mathrm{SB}(R',\theta') = \sum_{j=1}^N{\frac{L_j}{2\pi\sigma'^2_j q_j'} \exp
    \left[-\frac{1}{2\sigma'^2_j}\left(x'^2_j + \frac{y'^2_j}{q'^2_j} \right)\right]},
\end{equation}
with
\begin{equation}
  x'_j  = R' \sin(\theta'-\psi_j')
  \quad \mathrm{and} \quad
    y'_j  = R' \cos(\theta'-\psi_j'),
\end{equation}
where $(R',\theta')$ are polar coordinates on the sky-plane. The
Gaussian components have axial ratio $0\le q'_j\le1$, dispersion
$\sigma'_j$ along the major axis, and position angle $\psi_j'$,
measured counterclockwise from the $y'$-axis to the major axis of each
Gaussian. The misalignment angle $\psi_j'$ cannot be measured directly
as the position of the intrinsic $y'$-axis is not observable.  We
define
\begin{equation}
\label{eq:defpsi}
\psi^\prime_j=\psi+\Delta\psi^\prime_j
\qquad \mathrm{with }\qquad
\Delta\psi^\prime_1\equiv0,
\end{equation}
where $\Delta\psi'_j$ is the isophotal twist of each Gaussian, which
can be measured directly.

\subsection{From projected to intrinsic shape}
To determine the parameters of the Gaussians in eq.~(\ref{density}),
we fit the two-dimensional MGE model of eq.~(\ref{surf_twist}) to the
observed surface brightness. After assuming the space orientation
($\vartheta,\varphi,\psi$) of the galaxy, the relations between the
observed quantities $(\sigma'_j, q'_j, \psi'_j)$ and the intrinsic
ones $(\sigma_j, p_j, q_j)$ are given by Cappellari et al.\
(\nocite{2002MNRAS.333..400C} 2002; for a different formalism see Monnet et
al.\ \nocite{1992A&A...253..366M} 1992)
\begin{eqnarray}
1 \! -\! q_j^2
\hspace{-8pt}&=&\hspace{-8pt}
\displaystyle\frac{ \delta_j' \!\left[
    2\cos\!2\psi_j' \!+\! \sin\!2\psi_j'
    (\sec\vartheta\cot\varphi \!-\! \cos\vartheta\tan\varphi)
  \right] }{ 2\sin^2\!\vartheta \!\left[
    \delta_j' \cos \psi_j'
    (\cos\psi_j' \!+\! \cot\varphi\sec\vartheta\sin\psi_j') \!-\! 1
  \right] }, \label{MC00} \\
p_j^2 \!-\! q_j^2
\hspace{-8pt}&=&\hspace{-8pt}
\displaystyle\frac{ \delta_j' \!\left[
    2\cos\!2\psi_j' \!+\! \sin\!2\psi_j'
    (\cos\vartheta\cot\varphi \!-\! \sec\vartheta\tan\varphi)
  \right] }{ 2\sin^2\!\vartheta \!\left[
    \delta_j' \cos \psi_j'
    (\cos\psi_j' \!+\! \cot\varphi\sec\vartheta\sin\psi_j') \!-\! 1
  \right] }, \\
u^2_j
\hspace{-8pt}&=&\hspace{-8pt}
\displaystyle\frac{1}{q_j'}
\sqrt{p_j^2\cos^2\!\vartheta \!+\!
  q_j^2\sin^2\!\vartheta(p_j^2\cos^2\!\varphi+\sin^2\!\varphi)
},\label{MC01}
\end{eqnarray}
where $\delta_j'=1-q_j'^2$, and $u_j\equiv\sigma_j'/\sigma_j$, the
scale-length projection compression factor, which together with the
dimensionaless parameters $p_j$ and $q_j$ define the intrinsic
shape. The mathematical constraint $q_j>0$ and $p_j>0$ (or the
stronger and more physical constraint $q_j>0.2$ and $p_j>0.4$, which
gives the range of reasonable axis ratios for an elliptical galaxy,
\nocite{1981MNRAS.194..679B}{Binney} \& {de Vaucouleurs} 1981) implies that each Gaussian can be
deprojected only for a limited range of orientations \nocite{1992A&A...253..366M}(see
  also {Monnet} {et~al.} 1992). The orientations for which the whole
MGE model can be deprojected are located in the intersection of the
regions that are allowed by the individual Gaussian components.

\subsection{Constructing a realistic triaxial MGE}

The individual Gaussian components have no direct physical
significance, but their parameters provide constraints on other, more
important, quantities. We must therefore be careful that the MGE-model
does not result in spurious conditions on the physical properties of
the galaxy.  The allowed intrinsic orientation of the galaxy depends
on the axis ratios of the Gaussians in the superposition. It can be
easily verified numerically that the region in the space of the
rotation angles $(\vartheta,\varphi,\psi_j')$ for which a Gaussian
with a given observed flattening $q'_j$ can be deprojected increases
with $q'_j$: a round Gaussian ($q'_j=1$) can be deprojected for any
assumed intrinsic orientation, while an extremely flat one ($q'_j \ll
1$) can only be deprojected when the object is observed along one of
its principal planes.  Moreover, when a Gaussian has an photometric
twist $\Delta\psi_j'$ with respect to the other Gaussians in the MGE,
than the allowed deprojection region $(\vartheta,\varphi,\psi_j')$
becomes even smaller.

\begin{figure}
\plotone{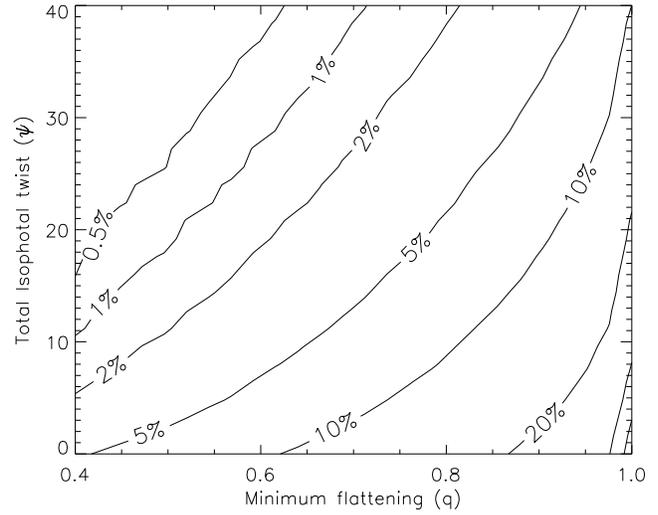}
\caption{\label{mgevolume} Contours of the deprojectable volume of an
  hypothetical MGE as a function of the observed isophotal twist and
  flattening.  The horizontal axis shows the minimum projected
  flattening and the vertical axis the maximum isophotal twist of all
  the Gaussians in the MGE. The labels denote the percentage of Euler
  angle space that can be deprojected.}
\end{figure}

The Gaussians in a given MGE-superposition generally have different
values of $q'_j$ and $\psi_j'$. This means that the MGE-model \emph{as
  a whole} can only be deprojected for angles that appear in the
intersection of the allowed individual regions
$(\vartheta,\varphi,\psi_j')$ of the deprojection of the individual
Gaussians. The largest deprojectable volume is obtained by maximizing
$\min\{q'_j\}$ and minimizing $\max\{|\Delta\psi_j'|\}$, while still
fitting the photometry within a certain accuracy \nocite{2002MNRAS.333..400C}(see
  also {Cappellari} 2002). This is verified numerically in
Fig.~\ref{mgevolume} which shows contours of the allowed volume
available for deprojection, for given minimum flattening and isophotal
twist of an MGE model.

The MGE models that are obtained in this way have, by construction,
the largest set of orientations for which a triaxial deprojection is
possible. For any given orientation, the roundest projection on the
sky corresponds to the roundest triaxial intrinsic density. This means
that this MGE model will be the roundest one that fits the
observations for any given intrinsic orientation of the galaxy. Very
boxy or very disky models are therefore excluded.

\subsection{Light to Mass}
\label{light2mass}

At this stage it is possible to add the contribution of invisible mass
to the gravitational potential of the model. This can be done by
creating two MGE models for the galaxy. One for the gravitating matter
and one for the visible light. The matter MGE is then used for the
calculation of the potential and the light MGE is used to reproduce
the intrinsic and observed light distribution.  To simulate a radial
$M/L$ profile one can construct a matter MGE by multiplying the
luminosity of each Gaussian of the light distribution with the desired
$(M/L)_j$ at that radius, see for example \nocite{2006ApJ...641..852V}{van den Bosch} {et~al.} (2006).
In this way, it is possible to construct a large range of potentials
or surface brightness distributions, as long as the matter and light
distributions can be represented by Gaussians. Alternatively one could
make $(M/L)$ in eq.~(\ref{density}) a function of $(x,y,z)$.

\subsection{Deprojection}
\label{deprojection}

The parameter range that has to be explored when fitting general
triaxial models to observations of elliptical galaxies is large. Two
axis ratios and three angles are needed to specify the intrinsic shape
and orientation of a triaxial ellipsoid, while only the projected
flattening $q'_j$ and the relative position angle $\Delta\psi'_j$ of
the projected major axis can be deduced from photometric
observations. As there is a relation between $q'_j$, the viewing
angles and the intrinsic shape parameters of the galaxy, the allowed
range of intrinsic shapes can be constrained to some degree, but a
large freedom remains. In some cases, additional information, such as
the value of the kinematic offset angle or the relative position of a
gas disk or dust lane, can provide further constraints
\nocite{1991ApJ...373..369B}(e.g., {Bertola} {et~al.} 1991).  However, unless two
perpendicular gas disks are observed, no unique intrinsic shape can be
deduced directly from the observations. The method to obtain triaxial
MGE mass models from surface brightness data that was described in the
above does not solve these problems.  However, it produces a range of
regular, searchable and well-behaved triaxial density distributions
that are consistent with the observed surface brightness, while being
easy to handle computationally.

The shape of the reconstructed potential used in our models is
directly related to the viewing angles by
eqs.~(\ref{MC00})-(\ref{MC01}). By changing the viewing angles the
potential of the model changes with it. However, the intrinsic shape
parameters are much more natural parameters than the viewing angles
($\vartheta,\varphi,\psi$), as they influence the appearance of the
orbits, and thus the kinematics they represent, much more directly.
For example, for an axisymmetric deprojection of the surface
brightness the angle $\varphi$ has no meaning, as rotating along the
symmetry axis does not change anything. Therefore we choose to study
the effects of the deprojection in terms of the intrinsic shape
parameters $(p,q,u)$, which can be computed from the viewing angles
($\vartheta,\varphi,\psi$) given the averaged flattening $q'$ of the
galaxy. The conversion from the intrinsic shape parameters to viewing
angles (which is the input for the models) is given by\footnote{The
  quantities $u^2$ and $u^2q'^2$ are recognized as the conical
  coordinates $\mu$ and $\nu$, with which the projected properties of
  a triaxial ellipsoid of axis ratios $p$ and $q$ can be evaluated in
  an elegant manner \nocite{1988MNRAS.231..285F}(e.g., {Franx} 1988).}

\begin{eqnarray}
\label{pqu2tpp}
  \cos^2\vartheta &=& 
  \frac{(u^2-q^2)(q'^2u^2-q^2)}{(1-q^2)(p^2-q^2)},
  \nonumber \\
  \tan^2\varphi &=&
  \frac{(u^2-p^2)(p^2-q'^2u^2)(1-q^2)}{(1-u^2)(1-q'^2u^2)(p^2-q^2)},
  \\ \nonumber
  \tan^2\psi &=&
  \frac{(1-q'^2u^2)(p^2-q'^2u^2)(u^2-q^2)}{(1-u^2)(u^2-p^2)(q'^2u^2-q^2)},
\end{eqnarray}
valid for $q\le p \le 1$, $q \le q'$ and $\max(q/q',p)\le u \le
\min(p/q',1)$. Four of the eight possible solutions are unphysical or
have $q>p$. The valid solutions are
\begin{equation}
  \label{validsoltpp}
  \begin{matrix}
    \{ & \vartheta     &,& -\varphi &,&  \psi & \}, \\
    \{ & \pi-\vartheta &,&  \varphi &,&  \psi & \}, \\
    \{ & \vartheta     &,&  \varphi &,& -\psi & \}, \\
    \{ & \pi-\vartheta &,& -\varphi &,& -\psi & \}.
  \end{matrix}
\end{equation}
They represent the same intrinsic shape only seen from the opposite
side and mirror images. They are thus identical and need not be
modelled separately.  However for Gaussians in the MGE with isophotal
twist ($|\Delta\psi_j'|>0$) the intrinsic shape of the models with
viewing angle $\psi$ and $-\psi$ are not the same, since the
$\Delta\psi_j'$ offset deprojects (eqs.~(\ref{MC00})-(\ref{MC01}))
them to a different ($p_j,q_j,u_j$), and thus a different intrinsic
shape. Hence, in the case of isophotal twist we have to consider one
solution from the first two lines in \eqref{validsoltpp}, and one from
the last two lines.

To convert from ($p,q,u$) to ($p_j,q_j,u_j$) one uses
eq.(~\ref{pqu2tpp}) and the averaged flattening $q'$ to go to
$(\vartheta,\varphi,\psi)$, and then eq.~(\ref{eq:defpsi}) (and the observed isophotal twists) to go to $\psi'_j$. From there one uses $q'_j$ and eqs.~(\ref{MC00})-(\ref{MC01}) to go to ($p_j,q_j,u_j$). 

To find the
best-fit intrinsic shape and corresponding viewing angles for an
observed galaxy the parameter space has to be searched effectively.
Since the models are computationally expensive the number of models
cannot be too large. The MGE parameterisation of the surface
brightness already excludes some viewing angles since their
deprojection is unphysical ($p<0.4$ or $q<0.4$). Especially an
isophotal twist reduces the allowed viewing angles. But also the
sampling in the intrinsic shape, instead of viewing angles, helps
reduce the number of models required, as this will avoid having models
with (nearly) the same intrinsic shape. Overall, for galaxies which
are mildly flattened approximately 100 distinct models are needed when
sampling $(p,q,u)$ in steps of 0.05.

\begin{table}
 
  \caption{Summary symbols introduced in Section~\ref{MGEsection}.}
  \begin{center}
  \begin{tabular}{@{\hspace{1pt}}l*{2}{@{\hspace{3pt}}l}@{\hspace{1pt}}}
    \hline\hline
    Symbol & Definition \\
    \hline
    $(x,y,z)$             & Intrinsic coordinate system \\
    $(x',y',z')$          & Projected coordinate system \\
    $(\vartheta,\varphi,\psi$)  & Viewing angles \\ 
    $(p,q,u)$             & Intrinsic shape parameters \\ 
    $q'$                  & Averaged projected flattening  \\
    $L_j,\sigma'_j,q'_j$  & Projected luminosity, dispersion, flattening \\
                          & \ \  of individual Gaussians \\
    $(p_j,q_j,u_j)$       & shape parameters of individual Gaussians\\
    $\sigma_j$            & Intrinsic dispersion of individual Gaussians\\
    $\psi'_j$             & Misalignment angle of individual Gaussians\\
    $\Delta\psi'_j$       & Isophotal twist of individual Gaussians\\
    \hline
  \end{tabular}
  \end{center}
 \label{tab:symbolsummary}
\end{table}

\subsection{Potential and accelerations}
\label{potacc}
The next step is to calculate the potential that corresponds to the
mass distribution of eq.~(\ref{density}). This is done by using the
classical \nocite{1969efe..book.....C}{Chandrasekhar} (1969) formula for the potential that
corresponds to a density stratified on similar concentric
ellipsoids. This results in \nocite{1994A&A...285..723E}({Emsellem} {et~al.} 1994)
\begin{equation}
  \label{pot}
  V(x,y,z) = -\sum\limits_{j=1}^{N} V_{0,j} 
  \int\limits_0^1\!d\tau F(x,y,z,\tau), 
\end{equation}
with
\begin{equation}
  V_{0,j} = (M/L) \; \sqrt{\frac{2}{\pi}} \frac{G L_j}{\sigma_j} 
\end{equation}
and
\begin{equation}
  F(x,y,z,\tau) = 
  \frac{\exp{\left[ 
        -\frac{\tau^2}{2\,\sigma^2_j}
        \left( x^2+ \frac{y^2}{1-\delta_j\tau^2} +
          \frac{z^2}{1-\epsilon_j\tau^2} \right) 
      \right]}}
  {\sqrt{(1-\delta_j\tau^2)(1-\epsilon_j\tau^2)}}, 
\end{equation}
where
\begin{equation}
  \delta_j=1-p^2_j 
  \quad \mathrm{and} \quad
  \epsilon_j=1-q^2_j.
\end{equation}
Here, $G$ is the gravitational constant and $M/L$ is the
mass-to-light ratio. Eq.~(\ref{pot}) has no simple analytic
expression and must be evaluated numerically. The integrand
is badly behaved in the central and outermost regions. It is therefore
more efficient to replace eq.~(\ref{pot}) by analytical
approximations in those regions.

The central density of each Gaussian can be expanded as
\begin{equation}
\rho_j(x,y,z)=\rho_{0,j}\sum\limits_{n=0}^{\infty}\alpha_n\,m^{2n},
\end{equation}
with $m^2=x^2+y^2/p^2+z^2/q^2$ and
\begin{equation}
  \alpha_n=\frac{1}{n!}\left(-\frac{1}{2\sigma^2_j}\right)^n.
\end{equation}
This expansion generates a potential [e.g. eq.~(29) of \nocite{1985MNRAS.215..713D}{de Zeeuw} \& {Lynden-Bell} 1985]
\begin{multline}
  \label{potexpand}
  V_j(x,y,z) = -\frac{V_{0,j}}{\sqrt{\epsilon_j}}
  \Biggl[ 
  F_j - \frac{1}{2\sigma^2_j} \left( A_{1,j}x^2+A_{2,j}y^2+A_{3,j}z^2 \right) 
  \Biggr. \\  
     +  \frac{1}{8\sigma^4_j} \left(     A_{11,j}x^4+A_{22,j}y^4+A_{33,j}z^4 \right. \\
  \Biggl. \left. + 2 A_{12,j}x^2 y^2+2 A_{13,j}x^2 z^2 + 2 A_{23,j} y^2 z^2 \right) + \dots
  \Biggr]
\end{multline}
The index symbols $A_{i}$ and $A_{il}$ are given in
\nocite{1969efe..book.....C}{Chandrasekhar} (1969). For a moderately triaxial model, the
expression~(\ref{potexpand}) differs less than $10^{-4}$ from the exact
potential for $r<0.1\,\sigma_j$, with $r^2 \equiv x^2+y^2+z^2$. A
higher-order Taylor expansion does not extend this limiting radius
significantly.

The potential outside $r>45\sigma_j$ can be approximated to within
$10^{-4}$ by the monopole term in a multipole expansion, which
corresponds to the potential of a central point mass with mass equal
to that of the Gaussian
\begin{equation}
  \label{dipole}
  V_j(x,y,z) = - (M/L) \; \frac{G\,L_j}{\sqrt{x^2+y^2+z^2}}.
\end{equation}
Higher-order multipole terms hardly extend the range of applicability.
Using equations (\ref{potexpand}) and (\ref{dipole}), numerical
integrations only have to be performed over the range $0.1\sigma_j
< r < 45\sigma_j$, which speeds up the orbit integration
significantly.

The contribution of a central supermassive black hole is represented
by a Plummer potential
\begin{equation}
  V_\bullet (x,y,z) = -\frac{G\,M_\bullet}{\sqrt{r^2_s+x^2+y^2+z^2}},
\end{equation}
in which $M_\bullet$ is the mass of the black hole and $r_s$ is a
softening length, which can be set to a non-zero value to prevent the
central potential to be infinite. In most applications, this smoothing
is used, and $r_s$ is chosen to be significantly smaller than the
smallest kinematic aperture. The black hole potential is added to
$V(x,y,z)$ from eq.~\eqref{pot} to obtain the total galaxy potential.
A separate dark halo potential can also be added at this stage, either
using the MGE (see \S~\ref{light2mass}) or another, specific,
expression.

The orbit integration is performed in Cartesian coordinates. The
stellar accelerations are given by the derivatives of equation
(\ref{pot}) with respect to $x,y$ and $z$. Similar to what is done for
the potential, the numerical calculation of the accelerations in the
central and outer regions of the model are replaced by, respectively,
a Taylor expansion and the dipole approximation. If we differentiate
the terms in eq.~(\ref{potexpand}), we obtain as first-order
approximations
\begin{eqnarray}
  a_{x,j} \hspace{-8pt}&=&\hspace{-8pt} 
  \frac{xV_{0,j}}{\sigma_j^2\sqrt{\epsilon_j}} \! 
  \left[ 
    \! A_{1,j} \!-\! \frac{1}{2\sigma_j^2}
    \left(A_{11,j}x^2\!+\!A_{12,j}y^2\!+\!A_{13,j}z^2 \right)
  \right] \!\! , 
  \nonumber \\ 
  a_{y,j} \hspace{-8pt}&=&\hspace{-8pt}
  \frac{yV_{0,j}}{\sigma_j^2\sqrt{\epsilon_j}} \!
  \left[
    \!A_{2,j} \!-\! \frac{1}{2\sigma_j^2}
    \left(A_{21,j}x^2\!+\!A_{22,j}y^2\!+\!A_{23,j}z^2 \right)
  \right] \!\! , 
  \\
  a_{z,j} \hspace{-8pt}&=&\hspace{-8pt}
  \frac{zV_{0,j}}{\sigma_j^2\sqrt{\epsilon_j}} \!
  \left[\!A_{3,j} \!-\! \frac{1}{2\sigma_j^2}
    \left(A_{31,j}x^2\!+\!A_{32,j}y^2\!+\!A_{33,j}z^2 \right)
  \right] \!\! ,
  \nonumber
\end{eqnarray}
where we have suppressed the dependence of the left side of the
equations on $(x,y,z)$. These expressions differ less than a factor
$10^{-4}$ from the exact accelerations inside $r<0.1\,\sigma_j$.  As
before, outside $r>45\,\sigma_j$ the accelerations can be approximated
to within $10^{-4}$ via the monopole term
\begin{equation} 
  a_{j,\xi} = (M/L) \; \frac{\xi\,G\,L_j}{\sqrt{(x^2+y^2+z^2)^3}},
  \qquad \xi=x,y,z.
\end{equation}
Similarly, the accelerations due to the black hole are given by
\begin{equation}
  \label{acc}
  a_{\bullet,\xi} = \frac{\xi\,G\,M_\bullet}{\sqrt{(r_s^2+x^2+y^2+z^2)^3}}.
  \qquad \xi=x,y,z.
\end{equation}
To make accurate and fast orbit integration possible, we interpolate
the total accelerations $(a_x,a_y,a_z)$ onto a three-dimensional polar
grid linearly in $[\log(r),\theta,\phi]$. For each grid point
$(r,\theta,\phi)$ we store
$[\log(-a_x/x),\log(-a_y/y),\log(-a_z/z)]$. We can then compute the
accelerations $(a_x,a_y,a_z)$ at point $(r,\theta,\phi)$ with
trilinear interpolation. After the interpolation grid has been
computed we ensure that the minimum relative accuracy is better than
$10^{-4}$.

\section{Orbits}
\label{orbits}

Schwarzschild's method tries to find a numerical representation of the
distribution function of a galaxy by assigning weights to a set of
orbits. To avoid any bias and to allow for the maximum degree of
freedom, the sample of orbits that the fitting routine can choose from
must be as general as possible and `representative' of the potential.
In this section, we describe how this is achieved. We first introduce
a triaxial Abel model from the companion paper (vdV07) that we use to
test our method. We then discuss the orbit structure in separable and
more general triaxial potentials. We continue with a description of
the orbital initial conditions, orbit integration and storage grids
that are used in our method.

\subsection{Separable test models}
\label{tests}

The Abel models with a separable potential from the companion paper
are a generalisation of the spherical Osipkov-Merritt models,
introduced by \nocite{1991MNRAS.252..606D}{Dejonghe} \& {Laurent} (1991) and extended by
\nocite{1999MNRAS.303..455M}{Mathieu} \& {Dejonghe} (1999). These models have a distribution function
(DF) that depends on three integrals of motions, contain a central
core, and allow for a large range of (triaxial) shapes.  The
observables of these models, including the LOSVD, can be calculated
efficiently and they can be used to generate test models that simulate
realistic wide-field imaging and integral-field spectrograph
kinematics of galaxies. These mock observations serve as input for the
triaxial Schwarzschild method presented in this paper.

We use the triaxial test model from \S4.3 of vdV07, which has an
isochrone St\"ackel potential. This model resembles a triaxial
$10^{11} M_{\odot}$ galaxy at 20 Mpc with a kinematically decoupled
component. We infer the potential from the MGE fit to the projected
(total) surface mass density. To obtain the luminous mass density, we
use a separate MGE that fits the surface brightness (assuming a
constant (stellar) mass-to-light ratio of $M/L=4$\,\MLsun). The
kinematics are constructed in such a way that they resemble \Sauron\ 
observations \nocite{2001MNRAS.326...23B}({Bacon} {et~al.} 2001).

We will use this test model to demonstrate our method.  More details
and tests of the recovery of global parameters are given in section
\S\ref{sec:test} of this paper, whereas tests of the recovery of the
internal structure and the DF can be found in the companion paper.

\subsection{Orbit structure}
\label{orbstruct}

\begin{figure}
\plotone{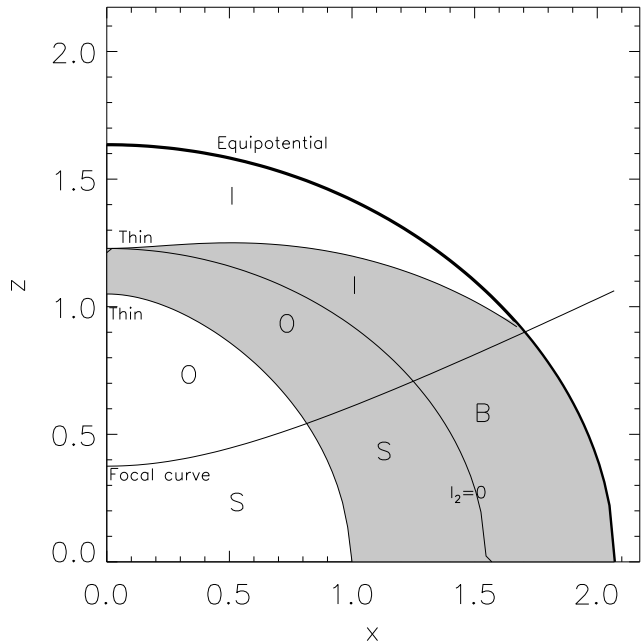}
\caption{\label{xz_one} The $(x,z)$-plane of a triaxial galaxy with a
  separable potential, for a value of the energy $E$ that is large
  enough that all orbit families appear.  The figure shows the
  equipotential that corresponds to $E$, the focal hyperbola, the
  curve at which $I_2=0$, and the location of the thin orbits. The
  regions where the different orbit families cross the $(x,z$)-plane
  perpendicularly are indicated: 'B' denotes box orbits, 'S'
  corresponds to short-axis tubes and 'I' and 'O' label inner and
  outer long-axis tubes. It can be seen that all tube orbits cross the
  $(x,z)$-plane perpendicularly in two points: once in the region
  outside the thin orbit curve and once inside. This means that the
  (grey) region between the thin orbit curves comprises all orbits
  just once, which is important for the orbital sampling
  ({Schwarzschild} 1993).}
\end{figure}

In a separable triaxial potential, all orbits are regular and conserve
three integrals of motion $E$, $I_2$ and $I_3$, which can be
calculated analytically. Four different orbit families exist: three
types of tube orbits, which avoid the center and are therefore
sometimes referred to as `centrophobic', and a set of orbits that can
cross the center, usually referred to as boxes or `centrophilic'
orbits \nocite{1973Kuzmin, 1985MNRAS.216..273D,
1987ApJ...321..113S}e.g., Kuzmin (1973); {de Zeeuw} (1985); {Statler} (1987).  These different orbit families conserve unique
combinations of these integrals and can therefore be linked to
distinct volumes in phase-space. Maybe even more remarkably, all four
orbit families in a separable potential cross the ($x,z$)-plane
perpendicularly in well-defined regions
(Fig.~\ref{xz_one}; Schwarzschild 1993).\nocite{1993ApJ...409..563S}
Similar to
axisymmetry, all tubes except the so-called thin orbits (in which the
inner and outer radial turning points coincide) cross the
$(x,z)$-plane perpendicularly twice. At a given energy, these points
are located in two distinct areas, separated by the line that connects
the points of the thin orbits. This line can be parameterized
analytically in a separable potential.

These properties are summarized in Fig.~\ref{xz_one}, where we have
used the isochrone separable potential of the triaxial Abel model. The
figure shows the $(x,z)$-plane for a value of the energy that is large
enough that all orbit families are populated. The thick outermost
curve is the equipotential at this energy, the inner- and outermost
decreasing curves inside the equipotential connect the points where
the thin orbits cross the $(x,z)$-plane perpendicularly, the
intermediate decreasing curve corresponds to $I_2=0$, and the rising
curve is the focal hyperbola. The four areas corresponding to the
different orbit families are also indicated (see \S5.4 of vdV07 for
further details).

This orbital structure depends crucially on the presence of a central
core and is (partially) destroyed by the addition of a supermassive
black hole and/or a central cusp \nocite{1985MNRAS.216..467G}({Gerhard} \& {Binney} 1985). Some orbits
in these non-separable potentials do not conserve global integrals of
motion other than the energy $E$ and may not all cross the
$(x,z)$-plane perpendicularly. The three types of tube orbits,
including the thin tubes, are still supported
\nocite{1993ApJ...409..563S}cf. {Schwarzschild} (1993). Most box orbits are transformed
into boxlets \nocite{1989ApJ...339..752M}({Miralda-Escud\'e} \& {Schwarzschild} 1989) and orbits that occupy
certain parts of phase space become chaotic. The amount of chaotic
motion and the radial range inside which it is present depends on the
central cusp slope (see \S\ref{blackhole}).

\subsection{Initial conditions}
\label{inicond}

The orbits in our models are more complicated than those in a
separable potential, as we use a more realistic MGE potential with a
supermassive black hole. Still, we use the properties of separable
models in our sampling of initial conditions. We sample the orbital
energy implicitly through a logarithmic grid in radius. When the model
has to reproduce observational data, it is important to sample the
orbital energy on a grid with a minimum radius that is at least an
order of magnitude smaller than the pixel size of the observations. In
the case of Hubble Space Telescope data, this typically corresponds to
$\sim 10^{-2}$ arcsec. The outer grid radius is determined by our
constraint that the grid must include $\geq 99.9$ per cent of the
mass.

Each of the grid radii $r_i$ is linked to an energy by calculating
the potential at $(x,y,z)=(r_i,0,0)$. The orbital initial conditions
are then sampled from a dense grid in the $(x,z)$-plane. Since most
orbits cross the $(x,z)$-plane perpendiculary twice above $z>0$ it is
not necessary to sample the whole plane. The double-countings are
avoided by finding the location of the thin orbit curves iteratively:
we launch orbits at different radii [keeping $\theta=\arctan(x/z)$
fixed] until the width of the orbit is minimal. This is similar to
what was done in the axisymmetric three-integral models, where all
orbits are short-axis tubes.

The starting points $(x, z)$ are selected from a linear open polar
grid $(R,\theta)$ in between the thin orbit and the equipotential (the
grey area in Fig.~\ref{xz_one}). The initial velocity in the $y$
direction is determined from $v^2_{y,0}=2[V(x_0,0,z_0)-E_i]$ and
$(v_x,v_z)=(0,0)$. This three-dimensional set $(E,R,\theta)$ of
starting conditions is commonly referred to as the `$(x,z)$ start
space' \nocite{1993ApJ...409..563S}({Schwarzschild} 1993). It is sufficient to launch
orbits in only one direction when the density (or another quantity
that is even in the velocity, such as the second moment) has to be
reproduced. When the velocity (and odd higher-order velocity moments
of the distribution function) is fitted in the model, the {\it
direction} of the orbital motion is also important. This information
could be taken into account directly by launching orbits in both the
positive and negative $y$-direction. However, the trajectories of the
pro- and retrograde orbits are identical, which means it is much more
efficient to include the counter-rotating orbits only at the fitting
stage by reversing the velocity sign appropriately. This is only valid
when figure rotation is ignored \nocite{1982ApJ...263..599S}cf. {Schwarzschild} (1982). 

\begin{figure*}
\plottwo{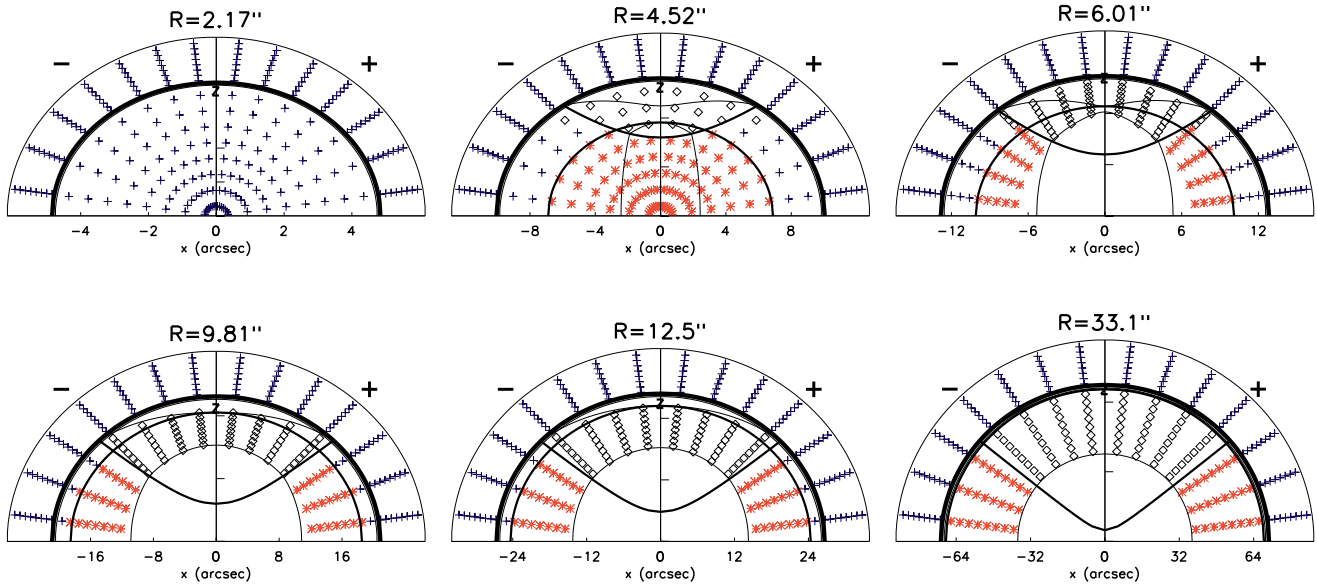}
\caption{\label{xz_startspace} {Representation of the ($x,z$) and the
    stationary start space and their symmetries for the triaxial Abel
    model from vdV07. The panels show the orbital starting points for
    increasing energies (denoted at the top), from an inner shell of
    the model (left top) to an outer shell (right bottom). The symbols
    represent the position of the orbits in the start spaces. The
    orbits in the \textit{inner right quarter} are in the ($x,z$)
    start space and the orbits placed in the \textit{outer right
      quarter} are in the stationary start space
    (\S\ref{inicond}). The thick black line represents the
    equipotential. (Compare to Fig.~\ref{xz_one}}.) The orbits in the
  \textit{inner left quarter} are the orbits from the ($x,z$) start
  space with reversed angular momentum and the orbits placed in the
  \textit{outer left quarter} is identical to the outer right quarter
  and are only drawn to make the panels symmetric. The symbols show
  the result of the orbit classification (based on angular momentum
  conservation, \S\ref{storage grids}): the crosses are box orbits,
  the stars correspond to short axis tubes and the diamonds correspond
  to (both types of) long axis tubes. We have also overplotted the
  analytical curves that separate the different type of orbits (see
  also Fig.~\ref{xz_one} and vdV07). The solid rising curve is the
  focal hyperbola, with above it the long axis tubes and below it the
  short axis tubes and boxes.  The crossing solid declining curve
  separates respectively between the inner and outer long axis tubes,
  and between the short axis tubes and boxes. The thin curves indicate
  the location of the corresponding thin tube orbits.}
\end{figure*}

Since boxes are essential for the support of the triaxial shape
\nocite{1979ApJ...232..236S, 1992ApJ...389...79H}({Schwarzschild} 1979; {Hunter} \& {de Zeeuw} 1992), a library with
relatively few of them cannot be expected to reproduce a triaxial mass
model. The ($x,z$) start space has few box orbits, especially at large
radii (see Fig.~\ref{xz_startspace}). To make sure that the orbit
library provides enough freedom in the outer parts of the model, we
add additional box orbits, like \nocite{1993ApJ...409..563S}{Schwarzschild} (1993). Box orbits
always touch the equipotential \nocite{1979ApJ...232..236S}({Schwarzschild} 1979). We
therefore sample start points on (successive) equipotential curves,
using linear steps in the two spherical angles $\theta$ and $\phi$. At
each combination of ($\theta,\phi$) and $E$, we use bisection to find
the corresponding value of $r_0$ that is located on the equipotential.
This three-dimensional set ($E,\theta,\phi$) of start conditions, the
`stationary start space' \nocite{1993ApJ...409..563S}({Schwarzschild} 1993), results in box
orbits or boxlets only. Tube orbits always conserve the sign of at
least one component of the angular momentum and therefore never reach
zero velocity. Since the direction of the orbits in this start space
is not important it is not necessary to add velocity mirrored copies
of them during the fit.

By design the set of energies $E$ and angles in $\theta$ in both start
spaces are identical, so that the orbits on the equipotential boundary
of the $(x,z)$ start space have obvious neighbours in stationary start
space. While not necessary, the size of the third dimension of the
start spaces is chosen to be the same for consistency.  Both sets of
orbits can be represented in a single figure, by graphically
connecting the corresponding starting spaces at the equipotential, as
shown in Fig.~\ref{xz_startspace}, where selected energies of the
triaxial Abel model (\S~\ref{tests}) are shown. In this figure we have
overplotted the same lines from Fig.~\ref{xz_one}, which shows that
our numerical scheme to locate the thin orbits indeed results in an
orbit sampling from the correct region. The stationary start space
intersects with the x-z start space at the equipotential. In the
figure all the orbits in the stationary start space that are nearest
to the equipotential are plotted just outside the
equipotential. Subsequent rows in the stationary startspace are
plotted radially outwards. A mirror image of the stationary start
space is also plotted for symmetry.

\subsection{Dithered orbit integration}

The orbits in the start space are integrated numerically and their
properties stored. The integration is done in Cartesian coordinates,
using an explicit Runga-Kutta method of order 5(4) \nocite{1993Hairer}(DOPRI5
routine by {Hairer}, {Norsett} \& {Wanner} 1993). With this method, the majority of the
orbits can be integrated with energy accuracies of better than one
part in $10^5$. This routine is capable of dense output, which enables
you to get a interpolated position and velocity at any time in current
timestep during the integration.

To ensure numerical precision the Runga-Kutta integrator uses more
steps where the orbital trajectory changes direction quickly. Since
this usually happens when the `star' is traveling with a high
velocity, the integrated time steps do not represent the time-averaged
path of the orbit. To make sure this is not a problem we use the dense
output of the integrator, to sample the orbit on equal time intervals,
ensuring that the orbits are properly time weighted.

Single orbits correspond to delta-functions in integral space, while
the distribution function of a (partially) phase-mixed galaxy is
expected to vary smoothly \nocite{1986MNRAS.219..285T}({Tremaine}, {Henon} \&  {Lynden-Bell} 1986). This limitation
may be reduced by combining nearby orbits \nocite{1988ApJ...327...82R,
1997ApJ...488..702R}({Richstone} \& {Tremaine} 1988; {Rix} {et~al.} 1997). Here we extend this method by `dithering' orbits
in all three dimensions in the initial starting space. We do this by
taking a bundle of $5^3$ orbits with different, but adjacent, initial
conditions and sum their observables. This method is also used in the
construction of axisymmetric models \nocite{2006MNRAS.366.1126C}see {Cappellari} {et~al.} (2006).

When calculating the starting spaces for the orbits we create more
starting points for the dithering. We enlarge the sampling
three-dimensional ($E,\theta,\phi$) start spaces five times in each
direction. This leads to 125 orbits per bundle. The odd number five
was chosen so that each bundle has a clearly defined central orbit
(see figure 6 in \nocite{2006MNRAS.366.1126C}{Cappellari} {et~al.} 2006). The orbital
properties of each of the orbits in each bundle are simpy co-added. As
an alternative, it is be possible to apply some form of (Gaussian)
weighting. In this way the orbit bundles could be made to overlap, but
the effects of this have not been studied.

Effectively, the model thus contains 125 times more orbits. The
dithering causes the orbital building blocks to be smoother,
eliminating aliasing effects, especially when modeling spatially
extended kinematic data. We found that this dithering is essential to
obtain smooth orbital observables and remove numerical noise, using a
limited amount of orbits.

\subsection{Storage grids and symmetries}
\label{storage grids}

For spherical galaxies, it is in principle sufficient to store the
orbital properties in one dimension, along a line. The
three-dimensional model can be reconstructed afterwards by
deprojecting the radial profile back onto the sphere. Similarly, axial
symmetry allows one to carry out the calculations in the meridional
plane only.
Revolution of the model around the intrinsic short axis returns the
three-dimensional intrinsic properties. As we restrict ourselves to
stationary, non-rotating galaxies that are symmetric in the three
principal planes, all orbital properties have to be calculated in only
one octant. The properties in the other octants follow by symmetry.

The density of every orbit in the library is stored on a spherical
grid in $(r_g,\theta_g,\phi_g)$ [$\theta_g=0$ corresponds to the short
axis and $(\theta_g=\pi/2,\phi_g=0)$ to the long axis]. The radial
sampling is logarithmic with the inner and outer boundary set to zero
and infinity. The angular grids $\theta_g$ and $\phi_g$ are sampled
linearly between $0$ and $\pi/2$. The grid has $N_{r_g}=15$, $
N_{\theta_g}=4$ and $N_{\phi_g}=5$. This leads to 20 bins per radius
and 300 bins in total, which is enough to ensure that the mass is
reproduced well and the model is self-consistent. When fitting the
model, the intrinsic mass grid is used as a constraint and is fitted
everywhere with an accuracy of 2\,\% (see~\S\ref{superposition}).

Similar to the intrinsic symmetries, the projected properties of
spherical galaxies are one-dimensional and those of axisymmetric
galaxies are symmetric in the projected axes. It is therefore
sufficient to store the projected properties of spherical galaxies in
one dimension and those of axisymmetric objects in one quadrant of the
sky. The projected properties of triaxial galaxies are at most
point-symmetric, with respect to the projected centre, which implies
that the model-data comparison must be done in one half (or more) of
the sky plane. 

To convert the intrinsic coordinates ($x$,$y$,$z$) to the projected
coordinates ($x'$,$y'$,$z'$) we use eqs.~(\ref{eqprojtointr}) and
(\ref{eqprojpsi}). After this step the PSF is included by randomly
perturbing the projected coordinates ($x'$,$y'$) with a
probability described by the MGE PSF, before being included into the
observational apertures (identical to \nocite{2006MNRAS.366.1126C}{Cappellari} {et~al.} 2006).
We use a three-dimensional rectangular storage grid in the projected
Cartesian coordinates $x'$ and $y'$ and the line-of-sight velocity $v$
in the the sky plane. The resolution and rotation of this grid is
adapted to the kinematical data that have to be reproduced.
Optionally, the observational apertures can be binned as a final step
to match any obervational binning.

\begin{table}
  \caption{
    The recipe that is used to mirror orbits in the three principal
    planes. Long-axis tubes are abbreviated by L-tube and short-axis tubes
    by S-tube.  \label{TableOne}} 
  \begin{center}
  \begin{tabular}{@{\hspace{1pt}}l*{3}{@{\hspace{3pt}}l}@{\hspace{1pt}}}
    \hline\hline
    Position & Box & L-tube & S-tube\\
    \hline
    $( x, y, z)$ & $( v_x, v_y, v_z)$ & $( v_x, v_y, v_z)$ & $( v_x, v_y, v_z)$\\
    $(-x, y, z)$ & $(-v_x, v_y, v_z)$ & $(-v_x, v_y, v_z)$ & $( v_x,-v_y, v_z)$\\
    $( x,-y, z)$ & $( v_x,-v_y, v_z)$ & $( v_x, v_y,-v_z)$ & $(-v_x, v_y, v_z)$\\
    $( x, y,-z)$ & $( v_x, v_y,-v_z)$ & $( v_x,-v_y, v_z)$ & $( v_x, v_y,-v_z)$\\
    $(-x,-y, z)$ & $(-v_x,-v_y, v_z)$ & $(-v_x, v_y,-v_z)$ & $(-v_x,-v_y, v_z)$\\
    $(-x, y,-z)$ & $(-v_x, v_y,-v_z)$ & $(-v_x,-v_y, v_z)$ & $( v_x,-v_y,-v_z)$\\
    $( x,-y,-z)$ & $( v_x,-v_y,-v_z)$ & $( v_x,-v_y,-v_z)$ & $(-v_x, v_y,-v_z)$\\
    $(-x,-y,-z)$ & $(-v_x,-v_y,-v_z)$ & $(-v_x,-v_y,-v_z)$ & $(-v_x,-v_y,-v_z)$\\
    \hline
  \end{tabular}
  \end{center}
\end{table}
Only orbits with the correct degree of symmetry can be used to
reproduce the density and potential. All orbits in a separable
potential are indeed eightfold symmetric, but this need not be the
case for resonant and irregular orbits in more general
potentials. These orbits can therefore not be used directly in the
reconstruction of the potential and density that we are interested
in. This does not mean that they are useless, as we can enforce the
required symmetries by apply a folding scheme to these orbits. Again,
this scheme is similar to what is done for axisymmetric potentials,
except that only orbits that are not symmetric with respect to the
$z=0$ plane have to be corrected in that case [e.g. 1:1 ($R,z$)
resonances, \nocite{1982ApJ...252..496R}{Richstone} 1982].

The folding scheme is based on the fact that a given asymmetric orbit
has up to seven mirror images that are obtained by reflection in the
principal planes (see the first column of Table~\ref{TableOne}). These
mirror images are also supported by the potential, but do not appear
in the library because we sample orbital initial conditions only from
one octant. All eight mirror orbits have identical properties and are
equally useful for the model. We may therefore add these eight orbits
to obtain an orbit that has three planes of symmetry. In practice,
this is done as follows. During orbit integration, the orbital weight
that corresponds to a given point $(x,y,z)$ is equally distributed
over the eight mirror points. The contributions to both the intrinsic
and projected density of all eight points are added up into one
orbital building block. In this way, orbits that are asymmetric are
included correctly, while orbits that already are eightfold symmetric
are simply sampled more densely.

More attention is required when calculating the kinematical
observables of the orbital building blocks. If we reflect the
velocities in the same way as the coordinates, the total orbital
building block has no net angular momentum. This is only correct for
box orbits, while tube orbits, which are essential when fitting to the
observed velocity field, conserve the sign of at least one component
of the angular momentum vector. Therefore, the sign of the angular
momentum of these orbits must be preserved also in the total orbital
building block. This is ensured by classifying the orbits on the basis
of their angular momentum properties. Box orbits oscillate in all
three directions, so that no components of the angular momentum are
conserved, while long-axis and short-axis tubes conserve the sign of
the angular momentum parallel to the long and short axis,
respectively.  This allows us to distinguish orbits by checking for
which angular momentum component(s), if any, the sign is conserved
during orbit integration.  In doing this, inner and outer long axis
tubes cannot be recognized separately. This is, however, not a problem
for the present application \nocite{1979ApJ...232..236S}(cf. {Schwarzschild} 1979). As can
be seen from Fig.~\ref{xz_startspace}, where we have plotted the
different orbital types with different symbols, the numerical
classification agrees with the analytical calculations.

We then apply the following scheme for reflections in the principal
planes (see Table~\ref{TableOne}). Box orbits: the average angular
momentum is zero, which allows us to reflect the velocity components
in exactly the same manner as the coordinates. Long-axis tube orbits:
the sign of the angular momentum around the long axis, $L_x=y\,v_z -
z\,v_y$, is conserved, which means that $L_x$ must be the same for all
eight mirror points. Short-axis tube orbits: the sign of $L_z=x\,v_y -
y\,v_x$ is conserved.

\subsection{The influence of a central mass concentration}
\label{blackhole}
Central mass concentrations have considerable influence on the orbital
structure of the galaxy as a whole and may induce chaotic behaviour.
In an axisymmetric potential, the non-integrable regions of
phase-space (usually referred to as the Arnold web) are not connected.
This means that the diffusion of chaotic orbits is limited and their
influence on the model is probably not significant, which justifies
the fact that chaotic orbits are not treated in a special manner in
axisymmetric models. Realistic triaxial potentials can support a much
larger fraction of chaotic orbits. The overall amount of chaotic
motion depends on the cusp slope and on the mass of the central mass
concentration \nocite{1985MNRAS.216..467G, 1996ApJ...460..136M,
1998ApJ...506..686V}({Gerhard} \& {Binney} 1985; {Merritt} \& {Fridman} 1996; {Valluri} \& {Merritt} 1998), but the different orbit families experience
fundamentally different effects, due to the central mass concentration

The box orbits that are started from the inner most equipotentials are
very difficult to integrate numerically. The central accelerations are
large and the time-steps that are required to conserve the integration
accuracy are correspondingly small, resulting in prohibitively long
orbital integration times. This effect can be reduced by using a
nonzero value for the softening length that was introduced in
\S\ref{potacc}. The DOPRI5 routine that we use to integrate the orbits
varies the time steps to match the desired accuracy. We found that
even orbits that pass close to the black hole can be integrated with
an accuracy of $10^{-5}$ in a reasonable time. The softening length
that we used is typically two orders of magnitudes smaller than the
radius of the sphere of influence of the black hole. This sphere of
influence is defined as $R_\bullet=GM/\sigma^2$, which is the radius
inside which the black hole potential dominates.

Test particles that are dropped from equipotentials at somewhat larger
distances from the black hole are scattered off their original (box)
orbits and may become chaotic \nocite{1985MNRAS.216..467G}({Gerhard} \& {Binney} 1985). The
trajectories of such particles can be described by a series of regular
segments and, given enough time, will fill most of the equipotential
that corresponds to the orbital energy. Since equipotentials are
rounder than equidensity curves and box orbits are the backbone of the
triaxial shape, this process may destroy the triaxial shape from the
inside out \nocite{2002ApJ...568L..89P}(e.g. {Poon} \& {Merritt} 2002).

Chaotic orbits are not time-independent, since their orbital densities
do not average out on physical time-scales. In principle, this means
that a Schwarzschild model with an orbit library that includes
irregular orbits is not stationary. However, evolutionary studies of
models that include chaotic orbits \nocite{1993ApJ...409..563S}({Schwarzschild} 1993) and
N-body simulations \nocite{1996ApJ...460..136M, 2002ApJ...567..817H}({Merritt} \& {Fridman} 1996; {Holley-Bockelmann} {et~al.} 2002)
display no dramatical shape-changes, even after very long times. This
means that a model with chaotic orbits may be stationary for as long
as a Hubble time and Schwarzschild solutions can be constructed also
for models that contain chaotic orbits.

The use of dithered orbital components (\S~\ref{implementation}) is
critical to create nearly time-independent models. In fact orbits
started from similar initial conditions can follow very different
trajectories. Because of the dithering single `sticky' chaotic orbits
do not play a major role, since orbits are always bundled with nearby
orbits.

The influence of a central mass concentration on tube orbits is
radically different. Low-energy tubes, which orbit at large radii,
never approach the central mass concentration close enough to be
significantly disturbed. Tubes that are launched from the principal
planes close to the radius of influence of the black hole turn into
precessing ellipses. Depending on the shape of the volume that the
ellipse eventually fills, they may be labeled as pyramid orbits
\nocite{2002ApJ...568L..89P}({Poon} \& {Merritt} 2002) or lens orbits \nocite{1999MNRAS.303..483S,
2000ApJ...542..143S}({Sridhar} \& {Touma} 1999; {Sambhus} \& {Sridhar} 2000). The precession rate of the ellipse is determined
by the ratio of the stellar mass that is enclosed by the orbit and the
central black hole mass. The integration time that is required for
convergence of the orbital properties is therefore inversely
proportional to the orbital radius.

\subsection{Number of orbits}

To summarize, we use two start spaces with three dimensions
$(E,\theta,R)$ and $(E,\theta,\phi)$, which are connected at the
equipotential boundary of every energy. The total number of orbits in
the fit, excluding dithering, is three times the number of orbits in
the one start space, as the $(x,z)$ start space is used twice and the
stationary start space once. The number of orbits is denoted as
$3\times N_E\times N_\theta \times N_R$. Because of the dithering each
effective orbit consists of 125 orbits, significantly smoothing the
orbital component. The total number of orbits necessary to make a
model is dependent on several factors: the number and spatial
distribution of the observed kinematics, the size of the galaxy model
and the shape of the potential.

The effect of the number of orbits can be studied by determining the
quality-of-fit $\chi^2$ of the model as a function of the number of
orbits. With increasing orbit numbers the $\chi^2$ decreases. When
enough orbits are present, the model does not improve any more and the
$\chi^2$ does not decrease anymore. In our test cases we find that the
model do not improve considerably when the orbit library consists of
2000 orbits or more. Self-consistent models with smaller orbit
libraries have significantly larger $\chi^2$, due to the fact that
there are not enough orbits to reproduce the mass, especially at
larger radii. We therefore decided to use an orbital sampling of 21
equipotential shells with $8\times7$ $(\theta,R)$ starting points
each, totaling 3528 $(\times125)$ orbits.

\section{Superposition and regularisation}
\label{superposition}

To make a model with the computed orbits we need to combine the orbits
in such a way that they fit the observations, while reproducing the
(intrinsic) mass distribution for self-consistency. Here we describe
the construction of the orbital superposition and a way to ensure that
our numerical solution is realistic.

\subsection{Finding the orbital weights}
\label{qp}

The model has two components that need to be fitted: The kinematic
observations and the (intrinsic and projected) mass distribution. The
kinematics are fitted using linearly superposed mass-weighted
Gauss-Hermite (GH) moments \nocite{1997ApJ...488..702R}({Rix} {et~al.} 1997). The fit is
done by combining the orbits linearly by assigning each orbit an
orbital weight ($\gamma_i$). These orbital weigths directly represent
the mass in each orbit and must therefore be positve ($\gamma_i \ge
$0).

The intrinsic mass grid (\S\ref{storage grids}) and the aperture
masses (the total amount of mass in each observed aperture: the zeroth
GH moment) must be added to the fit to ensure that the model is
self-consistent with the density in which the orbits were calculated.
Often this is done by including them in the fit as an `observable'
\nocite{1998ApJ...493..613V, 2004ApJ...602...66V}(e.g., {van der Marel} {et~al.} 1998; {Valluri} {et~al.} 2004). However
they are not actually observed and therefore it is difficult to assign
an error. To include them into the fit they are usually assigned a
hand-tuned fractional error so that the mass is reproduced well
without influencing the fit of the kinematics. Here we use a different
approach by including them as `constraints' with bounds in the fit \nocite{1988ApJ...327...82R}(similar to {Richstone} \& {Tremaine} 1988).
This means that the orbital superposition reproduces the intrinsic and
aperture masses to within 2\,\% at all times, while finding the
best-fitting kinematics. The total normalised mass of all the orbital
weights is fixed using an
equality constraint. 

The reason for including constraints is that the mass can almost
always be reproduced up to numerical percision
\nocite{1998ApJ...493..613V, 2002ApJ...568L..89P}({van der Marel} {et~al.} 1998; {Poon} \& {Merritt} 2002) and is thus not
relevant for finding the best-fit solution. We only want the mass to
be reproduced to within 2\,\%, because this is the typical accuracy of
(the MGE of) the observed surface brightness. Within these bounds the
solver allows the mass to vary to find the best-fitting kinematics in
the least squares sense.

We use the sparse quadratic progamming solver QPB from the GALAHAD
library \nocite{2003galahad}({Gould}, {Orban} \& {Toint} 2003) to make the superposition, as this
algorithm is capable of fitting the kinematics in the least squares
sense while satisfying mass constraints. This algorithm optimizes the
orbital weight $\boldsymbol{\gamma}$ in the least squares sense:
\begin{equation}
 \mathop {\min }\limits_{\boldsymbol{\gamma} \in R^n } 
 \left\| {\mathbf{A}\boldsymbol{\gamma} - \mathbf{b}} \right\|^2
\end{equation}
subject to the positivity constraint  
\begin{equation}
 \boldsymbol{\gamma} \ge 0,
\end{equation}
and the linear constraints,
\begin{equation}
 0.98\,\mathbf{p} \le \mathbf{M}\boldsymbol{\gamma} \le 1.02\,\mathbf{p}.
\end{equation}
Here $\mathbf{A}$ is the $m \times n$ projection matrix whose $n$
columns give the model contribution of every orbit to the $m$
kinematical observables $\mathbf{b}$. The matrix $\mathbf{M}$ is a
projection matrix giving the model contribution of the orbits to the
mass in the various apertures and $\mathbf{p}$ is the mass derived by
integrating the MGE model over the projected apertures and intrinsic
mass grid. The total number of constraints in the fit ($\mathbf{p}$)
is 300 (intrinsic mass grid) + 1 (total mass) + the number of
apertures (aperture masses).

The quality of the model is determined by measuring the discrepancy
between the model and the observations for different values of the
input parameters. This is done by calculating the $\chi^2$, defined as
\begin{eqnarray}
\chi^{2}=\sum_{i=1}^{N_d}\left(\frac{D^*_i-D_i}{\Delta D_i}\right)^{2},
\label{schi}
\end{eqnarray}
in which $N_d$ is the number of observables (the number of apertures
times the GH moments), $D_i$ is the observation for the $i$th
observable, $D^*_i$ is the model prediction and $\Delta D_i$ is the
uncertainty that is associated with this value (the observational
error).

\subsection{Regularisation scheme}
\label{nnls}

The quadratic programming problem to be solved is ill-conditioned in
most applications, due to (close to) degenerate orbits. As a
consequence, the orbital weight distribution for the solution with the
smallest $\chi^2$ may be a rapidly varying function, which is not
likely to be realistic \nocite{1993ApJ...413...79M,
2002MNRAS.331..959V}({Merritt} 1993; {Verolme} \& {de Zeeuw} 2002). This effect can be reduced by adding linear
regularisation equations to the problem
\nocite{1996MNRAS.283..149Z, 1997ApJ...488..702R}(e.g., {Zhao} 1996; {Rix} {et~al.} 1997), which is
also known as `damping' in the field of linear programming. Such
regularisation terms can be added to force the orbital weights towards
a smoother function by minimizing their higher-order derivatives.

Our two start spaces are sampled in three dimensions $(E,R,\theta)$
and $(E,\theta,\phi)$ and they are connected at the equipotential
boundary. The three dimensions of the $(x,z)$ start space roughly
correspond to the three integrals of motion (in a separable
potential).  We can thus enforce smoothness of our solution by adding
regularisation terms to our minimisation routine in each of the three
directions $(k,l,m)$ of our start space. We adopt the second order
finite differencing (Press et al. 1992) regularisation from
\nocite{1999ApJS..124..383C}{Cretton} {et~al.} (1999), so for each orbit we add three equations
to the array~\textbf{A} given above, and minimises them in a least
squares sense
\begin{equation}
\begin{matrix}
  \lambda(2\xi_{k}\gamma_{(k,l,m)} 
  - \xi_{k-1}\gamma_{(k-1,l,m)} 
  - \xi_{k+1}\gamma_{(k+1,l,m)})
  \!\!&\!\!=\!\!&\!\!0, \\
  \lambda(2\xi_{k}\gamma_{(k,l,m)} 
  - \xi_{k}\gamma_{(k,l-1,m)}  
  - \xi_{k}\gamma_{(k,l+1,m)})
  \!\!&\!\!=\!\!&\!\!0, \\
  \lambda(2\xi_{k}\gamma_{(k,l,m)} 
  - \xi_{k}\gamma_{(k,l,m-1)}
  - \xi_{k}\gamma_{(k,l,m+1)})
  \!\!&\!\!=\!\!&\!\!0,
\end{matrix}
\end{equation}
where $\gamma_{(k,l,m})$ represents the orbital weight at position
$(k,l,m)$ in the start space grid. The $\xi_k$ weights are added to
include the radial energy dependence of the model. It is estimated, a
priori, as the normalised mass enclosed by each radial shell in the
start space
\begin{equation}
\frac{1}{\xi_k}=\frac{1}{N_o\iiint\!\rho dxdydz} 
\iiint\limits ^{R_{(k+1)}}_{R_{(k-1)}}
\rho(x,y,z)dxdydz,
\end{equation}
where $N_o$ is the number of orbits. The regularisation error
$\lambda$ determines how much smoothing is perfomed. Increasing
$\lambda$ increases the amount of smoothing. The optimal value of this
$\lambda$ is usually time-consuming to determine \nocite{1999ApJS..124..383C}(see,
e.g., {Cretton} {et~al.} 1999). However, it has been shown elsewhere
that a theoretical axisymmetric galaxy with a two-integral
distribution function can be accurately reproduced by using this
approach \nocite{2002MNRAS.335..517V, 2004MNRAS.347L..31C}({Verolme} {et~al.} 2002; {Cretton} \& {Emsellem} 2004).
Applications of the axisymmetric Schwarzschild method have often used
a value of $1/\Delta\equiv\lambda=0.25$ as regularisation
\nocite{2002ApJ...578..787C, 2005MNRAS.357.1113K}({Cappellari} {et~al.} 2002; {Krajnovi{\'c}} {et~al.} 2005). As we will show in
the next section it is also acceptable for the triaxial method.

\subsection{Testing the regularisation}
\label{regtest}

\begin{figure*}
\plottwo{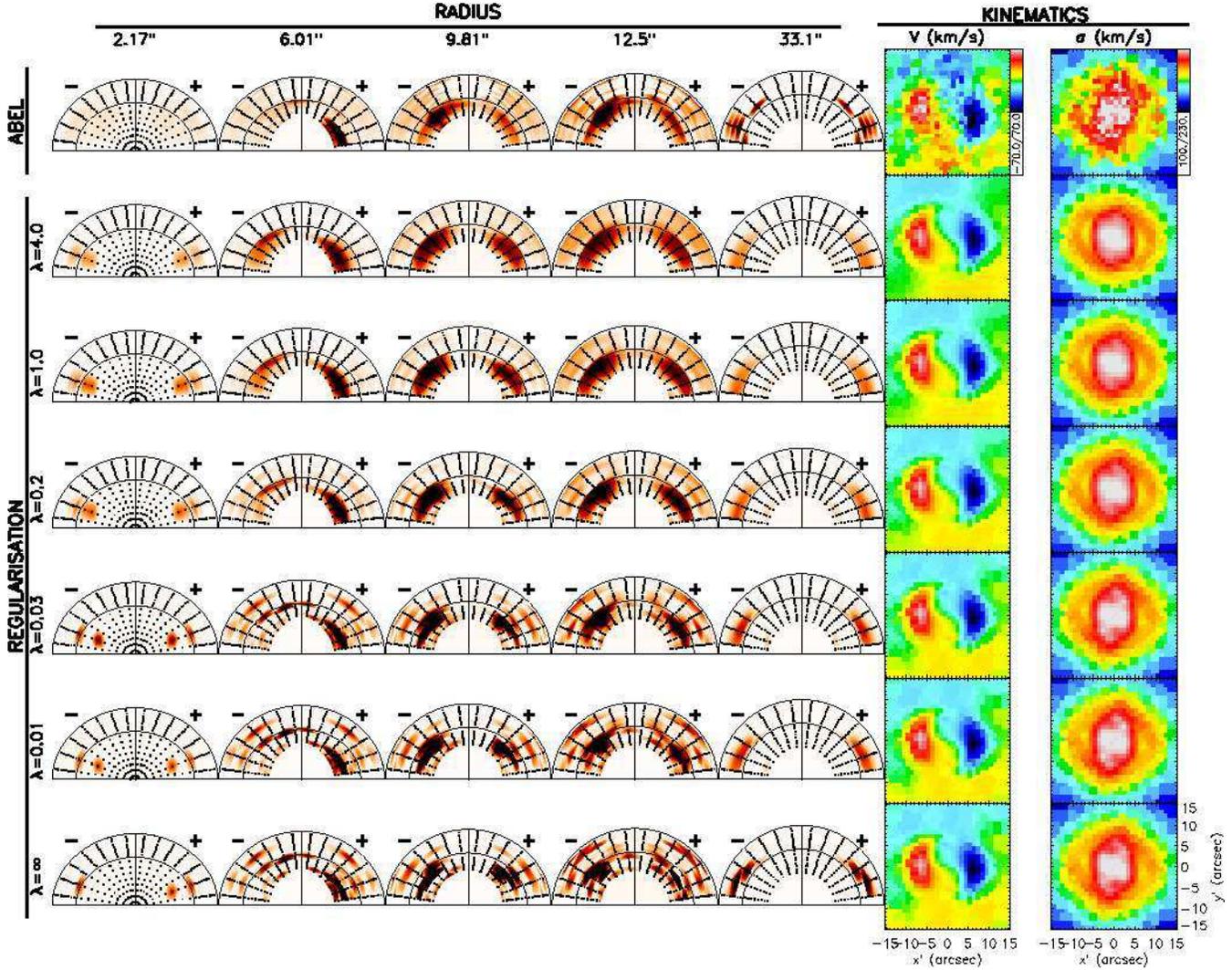}
\caption{\label{weights} The effect of regularisation on the
  models. The \textit{top row} shows the triaxial Abel from van de
  vdV07, while the \textit{other rows} show the best-fit Schwarzschild
  models with decreasing regularisation, from top to bottom. (For a
  detailed comparison of the DF see vdV07.) The \textit{left columns}
  show the orbital weights of the models in the same configuration as
  Fig.~\ref{xz_startspace}, whereas the \textit{two rightmost columns}
  show the velocity and velocity dispersion fields.}
\end{figure*}

In the companion paper the DF of the triaxial test model is compared
directly in terms of the integrals of motion (\S~5.4.3 in vdV07) and
we find that the recovery of the DF is consistent with the quality of
the input kinematics (Figs~12 and 13 in vdV07). The triaxial test model
is ideally suited to test regularisation. To do this we compare the
orbital mass weights directly. The top row in Fig.~\ref{weights}
displays the computed orbital weights and kinematics for the triaxial
Abel model from vdV07. The other rows show the orbital weights and
kinematics for the best-fit Schwarzschild model with decreasing
regularisation from top to bottom. The orbits at the radius of $2''$
and $33''$ are outside the range where they are constrained by the
kinematics, and as such the orbital weights for the Schwarzschild
models are not expected to compare well with those of the Abel model.

The distribution of (analytical) orbital weights for the Abel model is
smooth with some sharp peaks. The reconstructed orbital weights from
the Schwarzschild agree well with the analytical orbital weights,
except for high values of $\lambda$, which corresponds to strong
regularisation. The orbital weights of strongly regularised models are
distributed more smoothly, adjacent orbits receive similar weight and
the kinematics start to change. From Fig.~\ref{weights}, we see that
the kinematics is affected by the regularisation at
$\lambda\gtrsim0.2$, and thus the optimal regularisation in this case
has to be chosen to be $\lambda \lesssim 0.2$. This will give
satisfactory orbital weights, and kinematics that are consistent with
the observations. The comparison is best for a $\lambda \sim 0.1$

There are many reasons why the reconstructed orbital weights do not
exactly match the analytical orbital weights. Most of them are minor
numerical and discretisation effects. For example: (i) The
Schwarzschild method samples the galaxy with discrete orbits (computed
in the reconstructed potential), which are related to the DF via their
phase-space volume \nocite{1984ApJ...287..475V}({Vandervoort} 1984). The resulting orbital
mass weights are evaluated in an approximated and numerical way (\S5.4
in vdV07). (ii) Some symmetric orbits appear twice (or more often) in
the orbit library.  Without regularisation the (quadratic) solver
ignores the second identical copy of this orbit, as they do not
improve the fit.  However these orbits do get assigned weight in the
test case. A good example of this are the box orbits in the $(x,z)$
start space, as they are added twice to the fit [like all $(x,z)$
orbits].  These differences are only visible in this direct comparison
of the orbital weights.

To estimate the effect of regularisation on the kinematical fit more
quantitatively, we investigated the $\chi^2$ difference between the
models with different values of the regularisation $\lambda$. For the
best-fit model with 2370 kinematical observables the total $\chi^2$ is
2588. Adding regularisation does increase the $\chi^2$ of the model.
With $\lambda=0.01$ (little regularisation) it does not affect the fit
to the kinematics ($\Delta\chi^2\sim1$). When increasing $\lambda$
furter to 0.2 or even 4 (very strong regularisation), the
$\Delta\chi^2$ changes to $\sim200$ and $\sim 1000$, respectively. These
numbers reflect what one sees in Fig.~\ref{weights}: for $\lambda
\lesssim 0.2$ the kinematical fit does not change visibly, whereas for
higher $\lambda$ (stronger regularisation) the kinematical fit becomes
rapidly worse.

One other important question is whether the regularisation changes the
recovered input parameters, including the viewing angles,
mass-to-light raio, anisotropy and black hole mass. This is nearly
impossible to test with real galaxies, as their properties are
unknown. The Abel model has known parameters and was used to test the
recovery. We found that there is no difference in the best-fit
parameters when a regularisation of $\lambda=0.2$ was chosen. The
confidence intervals of the parameters do become smaller by using
regularisation. This is expected, as the added regularisation terms
decrease the freedom of the model and therefore increase the $\chi^2$.

A notable exception, that we do not test here, is the recovery of the
black hole mass. There are often few observables in the models near
the sphere of influence. The number of mass bins near the black hole
is extremely limited and the kinematical observations inside the
sphere of influence of the black hole is very limited, usually less
than 10 observables. In this scenario it is conceivable that
regularisation is needed, as the model might otherwise adapt the
orbital structure to be able to accomodate the black hole \nocite{2006MNRAS.373..425M}(see
also e.g., {Magorrian} 2006). Recovery of the black hole mass
using regularisation will be presented elsewhere.

\section{Tests on the triaxial Abel model}
\label{sec:test}

We test our method on the triaxial Abel model from the companion paper
vdV07, introduced in \S\ref{tests} and already used in
\S~\ref{regtest} above. Here, we outline further tests done on the
Abel model.

\subsection{Internal orbital structure and DF recovery}

In vdV07 the best-fit triaxial Schwarzschild model to the input
triaxial Abel is presented. The Schwarzschild method only uses the
information that can be observed in real galaxies, i.e. the
two-dimensional surface brightness and the two dimensional stellar
kinematics. The resulting best-fit model is an excellent fit and has a
(reduced) $\chi^2$ per degree-of-freedom of $\sim 1.1$.

Since Schwarzschild models only fit the projected observables it is
not obvious that these models can recover the three-integral DF and
the internal structure of the test model. By comparing the mass
weights of the Schwarzschild model to the DF of the test model, vdV07
demonstrate that both the internal orbital structure and DF are
recovered, with an accuracy similar to the typical (simulated) errors
on the kinematics.

\subsection{Recovering the global parameters}
\label{recoverglobalpar}

\begin{figure*}
\plotone{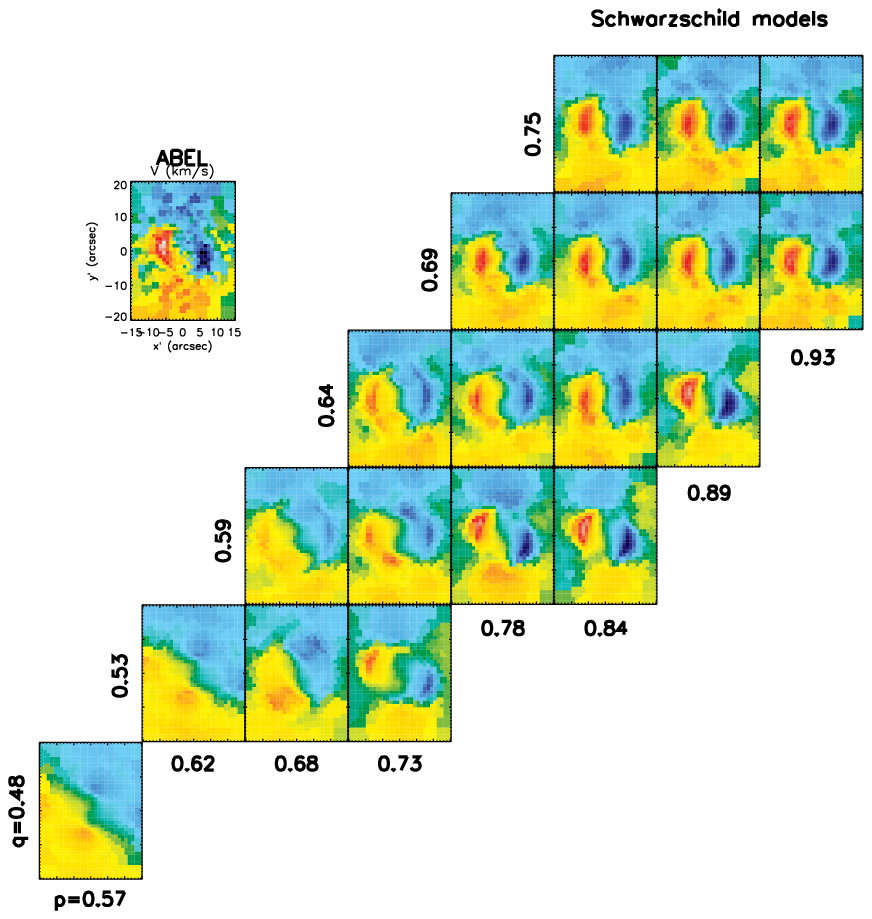}
\plotone{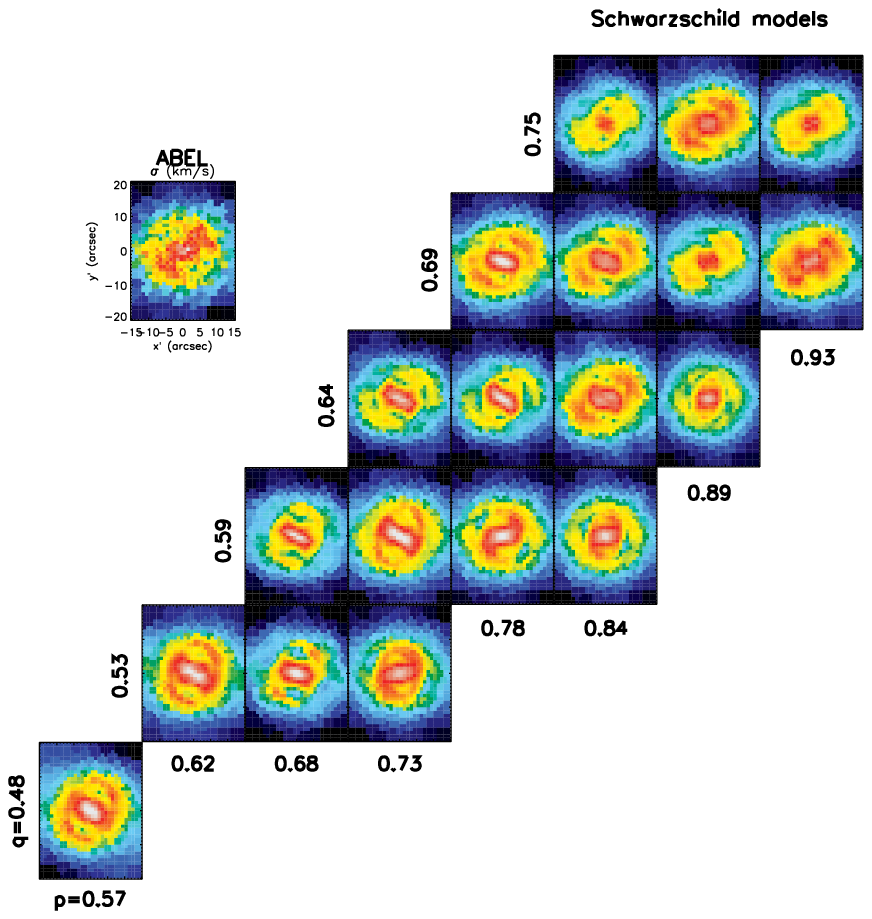}
\caption{\label{difpot} Schwarzschild models with different intrinsic
  shapes fitted to the observables of the triaxial Abel model from
  vdV07, reveal that the kinematics vary significantly by changing the axis ratios $p$ and $q$. The panels on the left show the velocity field of the Schwarzschild model for each value for $p$ and $q$, while the input velocity field from the Abel model is shown in the top left corner. The panels on the right shows the same for the velocity dispersion. The input model has axis ratios $(p,q)=(0.82, 0.67)$}
\end{figure*}

When constructing Schwarzschild models of the Abel model, we expect
the kinematics of the Schwarzschild model to vary when changing
intrinsic shape parameters $(p,q,u)$, and thus the viewing angles
$(\vartheta,\varphi,\psi)$. The most obvious are the change of the
zero-velocity curve as the projected axes change on the sky. Also, the
characteristics of the orbits in the model are dependent on the shape
of the potential. These effects are shown in Fig.~\ref{difpot}, by
showing models of the analytic test data in linear steps of 0.05 in
$p$ or $q$. The models become significantly worse when changing the
parameters away from the correct values. This shows that different
intrinsic shapes support different orbits and that one cannot expect a
model with the wrong potential to be able to fit the kinematics in all
cases.

\begin{figure}
\plotone{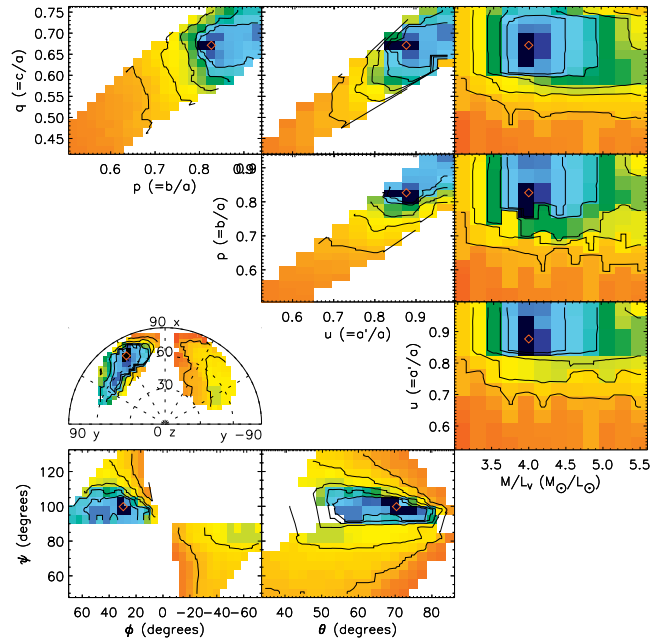}
\caption{\label{dchi2triax} Marginalized contours maps of the
  Schwarzschild models fitted to the observables of the triaxial Abel
  model for different intrinsic shapes. The contours denote 2, 4
  (thick line) 8 and 32 sigma confidence levels.  Areas for which the
  MGE cannot be deprojected are left blank. The six upper-left panel
  show the intrinsic shape parameters $(p,q,u)$ and mass-to-light
  ratio $M/L$; the three lower right panels show the viewing angles
  $(\vartheta,\varphi,\psi)$. The combination of $\vartheta$ and
  $\varphi$ is shown in a lambert equal area projection, seen down the
  north pole ($z$-axis). The $x$, $y$ and $z$ symbols give the
  location of views down those axis. The red diamond in each panel
  indicates the input parameters from the Abel model.}
\end{figure}

To search the global parameters we sample the parameter space, by
making linear steps of 0.1\,\MLsun\ in $M/L$, and 0.05 in $p$, $q$ and
$u$ (resulting in 100 different intrinsic shapes). For each
corresponding Schwarzschild model, the changes are quantified by the
goodness-of-fit parameter $\Delta\chi^2$. To visualize this
four-dimensional parameter space, we calculate for a pair of
parameters, say $p$ and $q$, the minimum $\Delta\chi^2$ as a function
of the remaining parameters, $u$ and $M/L$ in this case. The contour
plots of the resulting marginalized $\Delta\chi^2$ for all different
parameters for the Schwarzschild models fitted to the observables of
the Abel model are shown in Fig.~\ref{dchi2triax}. Since we sampled in
intrinsic shape and not in viewing direction, the viewing angle
sampling is not uniform.  In particular the very round models, which
are independent of $\phi$ are not represented properly. To this end,
we create a dense grid in $(p,q,u)$ and interpolate the $\chi^2$
linearly over this dense grid, resulting in the contour plots of
$(\vartheta,\varphi,\psi)$ in Fig.~\ref{dchi2triax}.

The input parameters for which the simulated observables of the Abel
model were obtained are $M/L=4$\,\MLsun\ and $(\vartheta,\varphi,\psi)
= (70^\circ,30^\circ,101^\circ)$. As outlined in
\S~\ref{deprojection}, the latter viewing angles convert to the
intrinsic shape parameters $(p,q,u) = (0.82,0.67,0.88)$, given the
average projected flattening $q'=0.76$ of (the MGE model of) the
surface brightness.  These input parameters are denoted by a red
diamond in the contour plots of Fig.~\ref{dchi2triax}. We find that
the input $M/L$ and $(p,q,u)$ (and hence also the viewing angles) of
the Abel model are accurately recovered, with a typical uncertainty of
10\,\% or less.

\nocite{2005MNRAS.357.1113K}{Krajnovi{\'c}} {et~al.} (2005) suggested that the recovery of the
inclination for axisymmetric models is degenerate, which seems in
conflict with our recovery of the intrinsic shape. However, the
kinematics of the Abel model have a significant feature, namely a
orthogonal decoupled core, and this makes it plausible that the
viewing angles are constrained quite strongly. We verified that for
galaxies with no such distinguished kinematic feature, e.g., in the
case of a (nearly) zero mean velocity map like for M87, the intrinsic
shape is not well constrained.

\section{Application to NGC~4365}
\label{ngc4365}

We now apply our method to NGC~4365, one of the prototypical galaxies
with a kinematically decoupled core (KDC). It is a giant E3 elliptical
and it was one of the first objects in which minor axis rotation was
discovered \nocite{1988A&A...195L...5W,1994MNRAS.269..785B}({Wagner}, {Bender} \&  {Moellenhoff} 1988; {Bender}, {Saglia}, \&  {Gerhard} 1994). The
peculiar velocity structure of this galaxy was partially unraveled by
multiple long-slit observations \nocite{1995A&A...298..405S}({Surma} \& {Bender} 1995), but the
full two-dimensional kinematical structure was only revealed with the
integral-field spectrograph \Sauron\! \,\nocite{2001ApJ...548L..33D}({Davies} {et~al.} 2001).

Kinematically decoupled cores can be the result of a merger event, but
can also occur when the galaxy is triaxial and supports different
orbital types in the core and main body \nocite{1991AJ....102..882S}({Statler} 1991).
\nocite{2001ApJ...548L..33D}{Davies} {et~al.} (2001) studied the first option and investigated
the link between the kinematics and the line-strength distribution of
NGC~4365. They found that the core and the main body are of similar
age and that any mergers that led to the formation of the KDC must
have occurred at least 12 Gyr ago, as otherwise younger stellar
populations would have been detected. The orbital structure that
supports the KDC and the main body cannot be observed directly and
must be inferred from dynamical models. \nocite{2004MNRAS.353....1S}{Statler} {et~al.} (2004)
studied the viewing angles and triaxiality of the system using an
approach developed by \nocite{1994ApJ...425..458S}{Statler} (1994a), which uses bayesian
analysis to fit analytic solutions of the continuity equation to an
observed velocity field.  They found NGC~4365 to be strongly triaxial
and seen almost along the long axis. The triaxial Schwarzschild method
that was presented in the previous sections allows us to build
comprehensive dynamical models of this galaxy and investigate its
intrinsic structure.

\subsection{Observations}

NGC~4365 was observed with \Sauron\ on the nights of March 29 and 30,
2000, for two different pointings, with an overlap in the central
region. The exposures were combined and rebinned into a map with a
slightly better spatial sampling (0.8 arcsec, compared to 0.94 arcsec
for the individual lenslets) and a coverage of 33 $\times$ 63 arcsec.
\nocite{2001ApJ...548L..33D}{Davies} {et~al.} (2001) give a full description of the
observations.

To increase the signal-to-noise ($S/N$) ratio to sufficient levels for
accurate determination of the kinematics, the datacube was spatially
binned into 964 non-overlapping bins using the two-dimensional Voronoi
binning of \nocite{2003MNRAS.342..345C}{Cappellari} \& {Copin} (2003). A minimum $S/N$ of 100 per
spectral element was imposed. However, many of the spectra have a much
higher $S/N$ value (up to $\sim300$), and more than one quarter of the
spectral elements remain unbinned. The stellar kinematics where
extracted using the penalised pixel-fitting method (pPXF) of
\nocite{2004PASP..116..138C}{Cappellari} \& {Emsellem} (2004). For every Voronoi bin we extracted the
velocity $V$, velocity dispersion $\sigma$ and the higher-order GH
moments $h_3$ and $h_4$ of the stellar LOSVD.

The \Sauron\ spectra have very high $S/N$ so that the LOSVD can be
reliably extracted from the data, however care has to be taken to
minimize the effect of template mismatch, which dominates the error
budget in the bright, high-$S/N$ central regions of the galaxy. To
this end an accurate template was determined during the pPXF fit using
the $\sim 1000$ stars of the MILES library \nocite{2006miles}({S{\'a}nchez-Bl{\'a}zquez} {et~al.} 2006), which
span a large range of stellar atmospheric parameters.  Out of the
MILES stars, only 14 are selected by pPXF to provide an accurate match
to the observed average galaxy spectrum, with an rms scatter in the
residuals of only 0.17\,\%. From the observed residuals and Fig.~B3 of
\nocite{2004MNRAS.352..721E}{Emsellem} {et~al.} (2004), we infer an upper limit of $\la0.02$ on
the systematic error of the GH moments, due to any remaining template
mismatch.

\begin{figure*}
  \plottwo{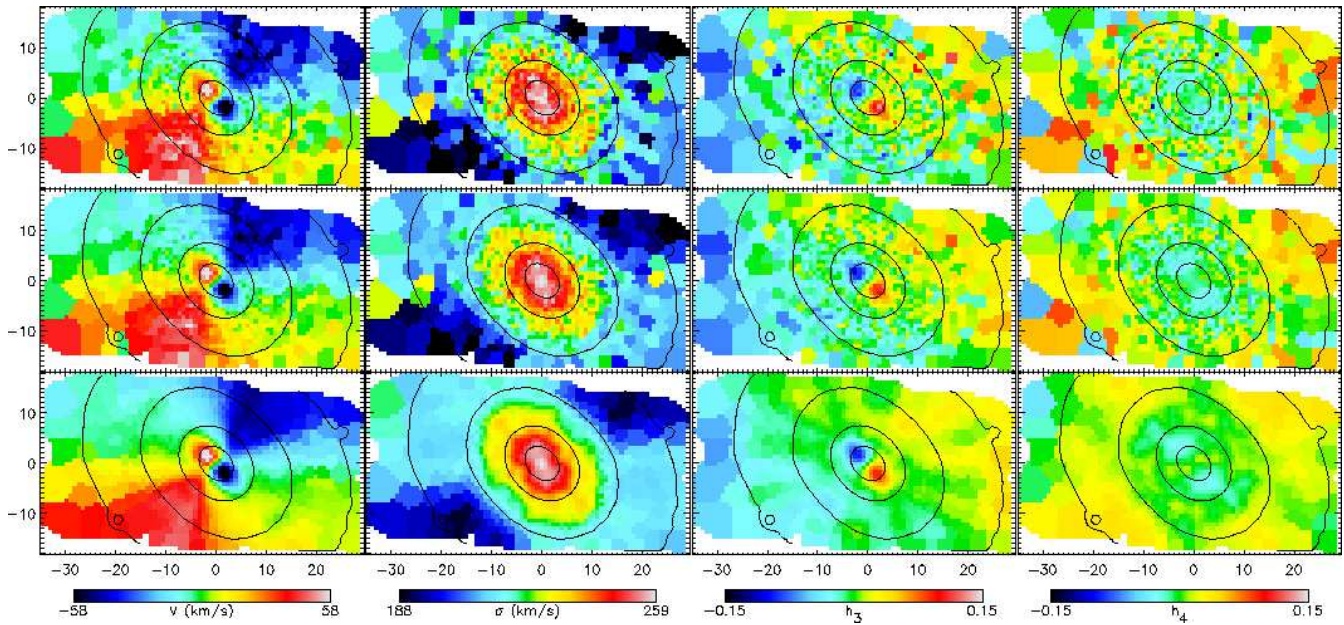}
  \caption{SAURON observations of NGC~4365. {\em Top panels:}
    from left to right: the mean velocity, velocity dispersion and GH
    moments $h_3$ and $h_4$ of NGC~4365, as observed with the
    integral-field spectrograph \Sauron. The pixels scale of the
    observations is 0\farcs8. {\em Middle panels:} point symmetrized
    kinematics with respect to the galaxy center. Non symmetric
    deviations cannot be reproduced by a triaxial model anyway and the
    symmetrization guides the eye. {\em Bottom panels:} kinematic maps
    of the best-fit Schwarzschild model, obtained by adding the
    weighted contributions of the best-fitting set of orbits.  The
    same color levels are used for both data and model.}
\label{SAURONmaps}
\end{figure*}

The re-reduced kinematics, shown in Fig.~\ref{SAURONmaps}, are a
significant improvement over the kinematics shown in
\nocite{2001ApJ...548L..33D}{Davies} {et~al.} (2001). They show a core in the inner $\sim6$
arcsec that rotates around the minor axis. At larger radii the stars
rotate around an axis offset by 82\dgr, which is evidence that the
system is intrinsically triaxial. The peak mean streaming velocities are 55
km\,s$^{-1}$. The dispersion peaks at a value of $\sim260$
km\,s$^{-1}$.

\subsection{Mass model}

\begin{figure}
  \plotone{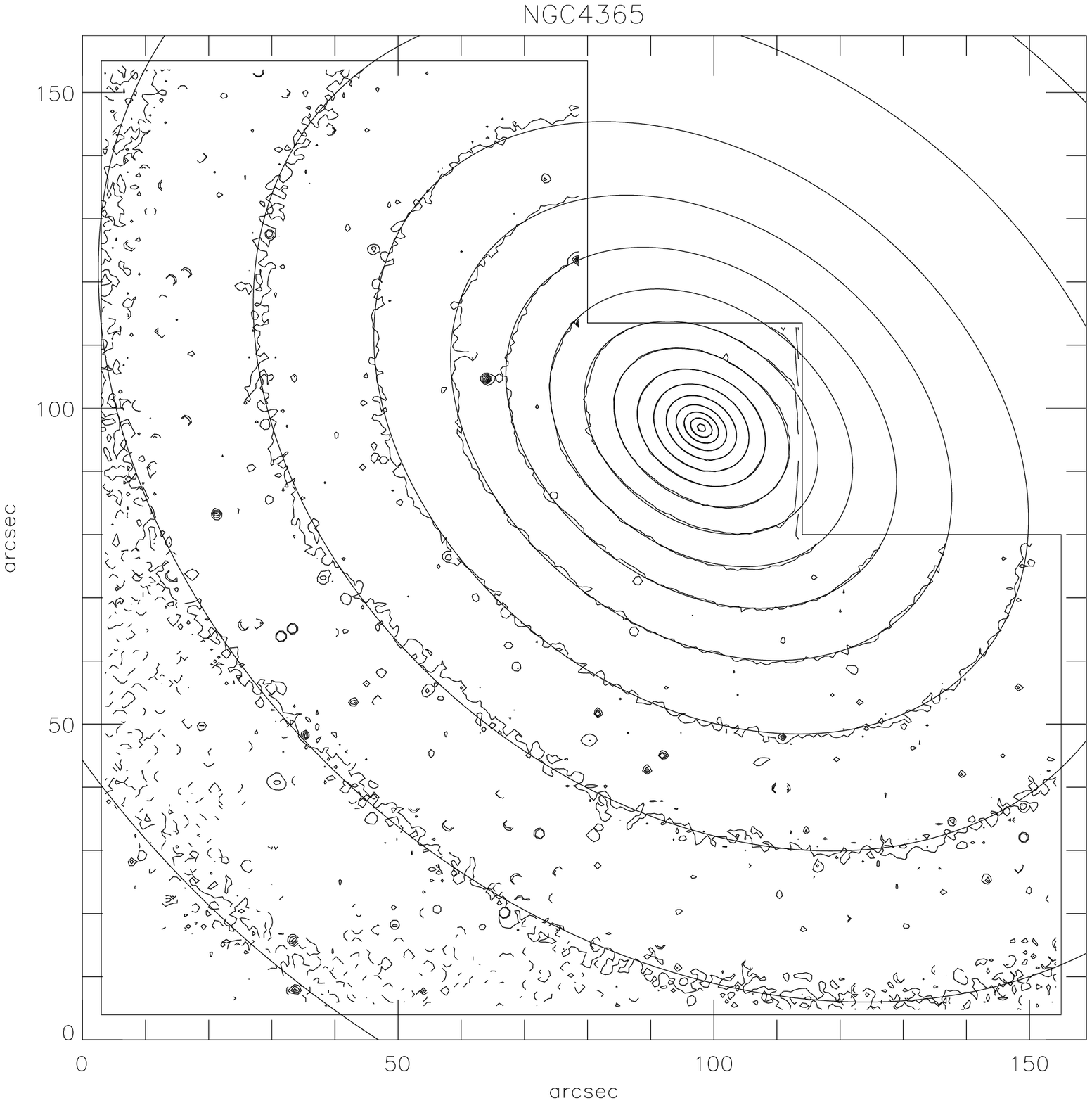}
  \plotone{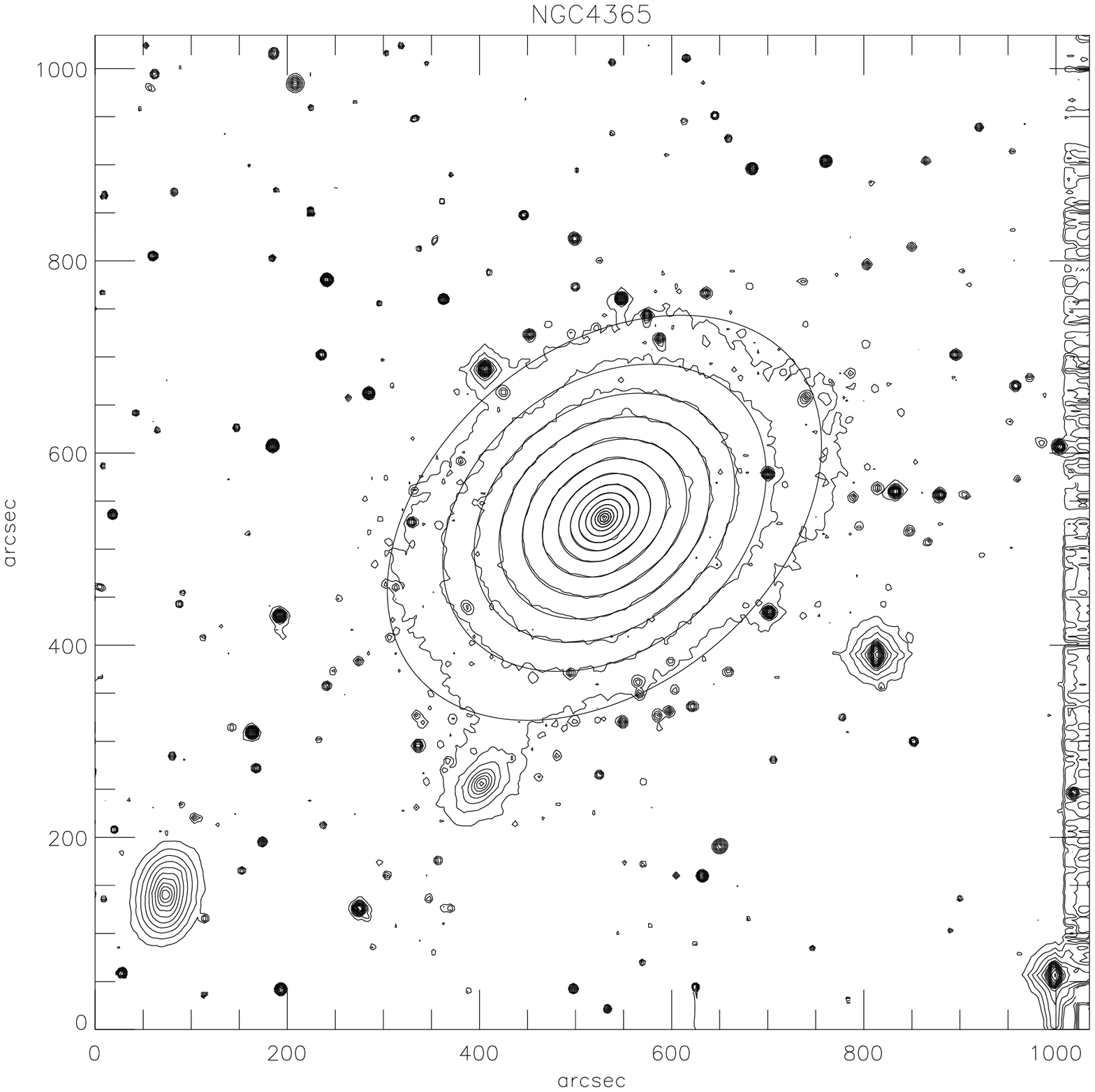}
  \caption{{\em Top:} The contours of \texttt{HST}/WFPC2/F814W image
    of NGC~4365, overplotted with (smooth) contours of the
    best-fitting MGE. {\em Bottom:} The contours of the ground based
    image of NGC~4365 obtained with the 1.3m McGraw-Hill telescope and
    the best-fitting MGE.}
  \label{MGEfit} 
\end{figure}

We used an \texttt{HST}/WFPC2/F814W image and a ground based image of
NGC~4365 obtained with the 1.3m McGraw-Hill at the MDM observatory
(from Falc\'on-Barroso et al., in prep.)  to make an MGE (mass) model,
using the software by \nocite{2002MNRAS.333..400C}{Cappellari} (2002). We ensure that the
model is the roundest that is consistent with the observations.  This
is done by setting a lower limit to the allowed projected flattening
of 0.67 and an upper limit of 4.5$^\circ$ on the difference in
position angle between the individual Gaussians.
The modest difference between the RMS-error of the free model (0.99
per cent) and the constrained MGE-model (1.02 per cent) suggests that
these constraints do not lead to systematic errors in the mass model.
The parameters of the MGE-model are given in Table~\ref{MGEtable}, the
surface brightness map and the MGE model fit are shown in Fig.
\ref{MGEfit}.

\begin{table}
\caption{The parameters of the eleven Gaussians in the MGE fit to the
  combined \texttt{HST}/WFPC2/F814W and the ground-based image of
  NGC~4365. The colums give for each Gaussian respectively its number
  $j$, amplitude $\mathrm{SB}_0=L/(2\pi\sigma'^2q')$, dispersion
  $\sigma'$, projected flattening $q'$, and position angle offset
  $\Delta\psi'$, as defined in eq.~\eqref{surf_twist}.}
\label{MGEtable}
\begin{center}
\begin{tabular}{l c c c c}
j  & $\log\mathrm{SB}_0$ (\Lsunpcsq) & $\log\sigma'$ (arcsec) & $q'$ &
$\Delta\psi'$ (\dgr)\\
\hline 
1   &    3.424  &    -1.024  &     0.800 & 0.0 \\
2   &    3.319  &    -0.727  &     0.800 & 0.0 \\
3   &    3.238  &    -0.320  &     0.800 & 0.0 \\
4   &    3.435  &    -0.027  &     0.670 & 0.0 \\
5   &    3.820  &     0.138  &     0.709 & 0.5 \\
6   &    3.740  &     0.402  &     0.698 & 0.8 \\
7   &    3.576  &     0.648  &     0.798 & 0.0 \\
8   &    3.106  &     0.955  &     0.737 & 0.0 \\
9   &    2.874  &     1.224  &     0.739 & 0.0 \\
10  &    2.400  &     1.499  &     0.741 & 3.5 \\
11  &    2.122  &     1.833  &     0.775 & 3.6 \\
12  &    1.329  &     2.362  &     0.670 & 4.5 \\
\hline
\end{tabular}
\end{center}
\end{table}

\subsection{Dynamical models}
\label{dynmodels}

We calculate triaxial Schwarzschild models using orbit libraries of
$3\times1176$ orbits, 2352 of which are started in the $(x,z)$-plane,
the remaining are dropped from the equipotential. We assume a distance
of 23 Mpc for NGC~4365 \nocite{2005ApJ...625..121M}({Mei} {et~al.} 2005). The assumed
distance does not influence our conclusions about the internal
structure of the galaxy, but lengths and masses scale linearly with
the distance, while mass-to-light ratios are inversely proportional to
the distance.

\begin{figure}
\plotone{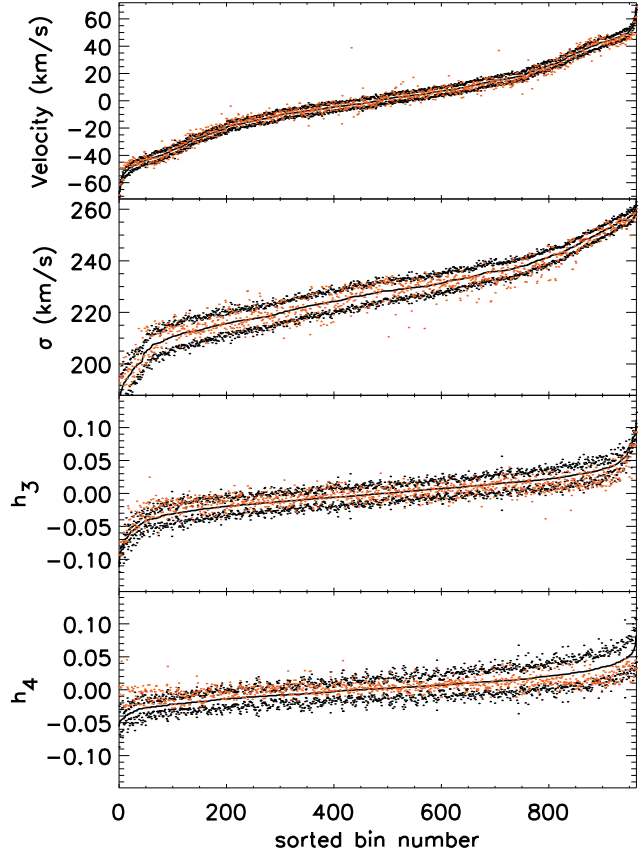}
\caption{The panels show the kinematics of the
  best-fitting model of NGC~4365. The line (with error bounds, shown
  as black dots) are the (sorted) observations, while the red dots
  show the model predictions.}
\label{cross} 
\end{figure}

\begin{figure}
  \plotone{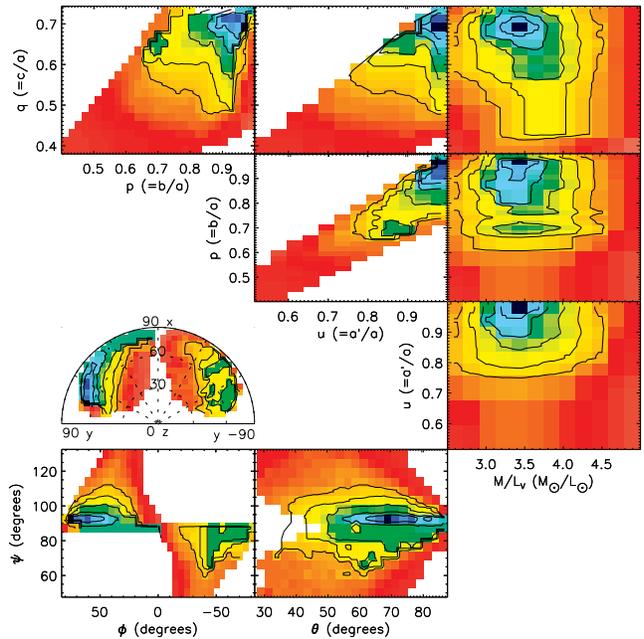}
\caption{Marginalized contours maps of the models of NGC~4365 for
  different mass-to-light ratios, shape parameters and (corresponding)
  viewing angles. The layout is identical to that of
  Fig.~\ref{dchi2triax}.  See \S~\ref{bestfitmodel} for discussion.}
\label{pqugal}
\end{figure}

A given triaxial model is determined by the mass-to-light ratio $M/L$,
the shape parameters $(p,q,u)$ --- or equivalently, the viewing angles
$(\vartheta,\varphi,\psi)$ --- and the mass $M_\bullet$ of the central
black hole.  We fix the latter to $M_\bullet = 3.6\times10^8$\,\Msun,
consistent with the black hole-sigma relation
\nocite{2002ApJ...574..740T}({Tremaine} {et~al.} 2002), as the \Sauron\ observations do not have
enough spatial resolution to resolve the radius of influence of the
black hole and therefore can not constrain the mass directly.
Given a typical flattening of $q'=0.74$ of the Gaussians in the MGE
model (Table~\ref{MGEtable}), we sample $(p,q,u)$ linearly in steps of
0.06. This results in 96 different intrinsic shapes which we combine
with $M/L$ values sampled linearly in 11 steps, from 3.0 to 5.0 in
solar units. This results in a total of 1056 Schwarzschild models, for
each of which we compute the goodness-of-fit to the observations via
$\chi^2$. The resulting models show a smooth gradient in $\chi^2$ as
can be seen in the $\Delta\chi^2$ contours in Fig.~\ref{pqugal}. As
before (see \S~\ref{recoverglobalpar}), to avoid an incomplete
sampling in viewing angle space, we oversample in $(p,q,u)$ before
computing the correspinding $\Delta\chi^2$ contours in
$(\vartheta,\varphi,\psi)$.
 
\subsection{Best-fit model}
\label{bestfitmodel}

The best-fit model (the model with the lowest $\chi^2$) is shown in
Fig.~\ref{SAURONmaps}. To be able to estimate the quality of the fit,
Fig.~\ref{cross} shows the residuals per velocity moment. Each panel
shows a moment sorted in order of increasing value. The observed
moment is indicated by the black line, with errors shown as black
dots. The red dots represent the corresponding values from the
best-fit model, which is in good agreement with the data. The
corresponding $\chi^2=4295$, which implies a $\chi^2$ per
degree-of-freedom (DOF) of 1.1. Statistically the standard deviation
of the $\chi^2$ is $\sqrt{2(N_\mathrm{obs}-N_\mathrm{par})}$. Given
that we have 3856 observables in our model and less than 10 parameters
($N_\mathrm{par}$), the expected scatter in the $\chi^2$ is much
larger than the customary $\Delta\chi^2=1$ criterion. We therefore
treat $\Delta\chi^2<\sqrt{2N_\mathrm{obs}}$ as a one sigma (68\%)
result.  The best-fit parameters are then $M/L=(3.5\pm0.2)$\,\MLsun\ 
(in $I$-band) and the intrinsic shape parameters $(p,q,u)
=(0.97^{+0.03}_{-0.05},0.70^{+0.03}_{-0.03},0.99^{+0.01}_{-0.04})$.

The best-fit $M/L$ is just consistent with the value of
$(4.3\pm0.4)$\,\MLsun\ predicted by the $M/L$-$\sigma$ relation
derived from \emph{axisymmetric} models by \nocite{2006MNRAS.366.1126C}{Cappellari} {et~al.} (2006),
using a $\sigma_e=231$\,\kms\ for this galaxy\footnote{The velocity
  dispersion $\sigma_e$ is derived from the \Sauron\ kinematics by
  luminosity-weighting all the spectra within one effective
  (half-light) radius and fitting a single Gaussian LOSVD.}.  It is
interesting to see that the best-fitting $M/L$ does not vary
significantly with any of the other model parameters as can be seen in
the $\Delta\chi^2$ contour plots in Figs.~\ref{pqugal}. For example,
even for models with a non-optimal value of $p,q$ or $u$, still the
best-fit $M/L\sim3.5$\,\MLsun. The total mass of the model, obtained
by converting the total luminosity to mass using the best-fit $M/L$
value is $4.8\times10^{11}$\,\Msun.

\subsection{Intrinsic shape}

\begin{figure}
\plotone{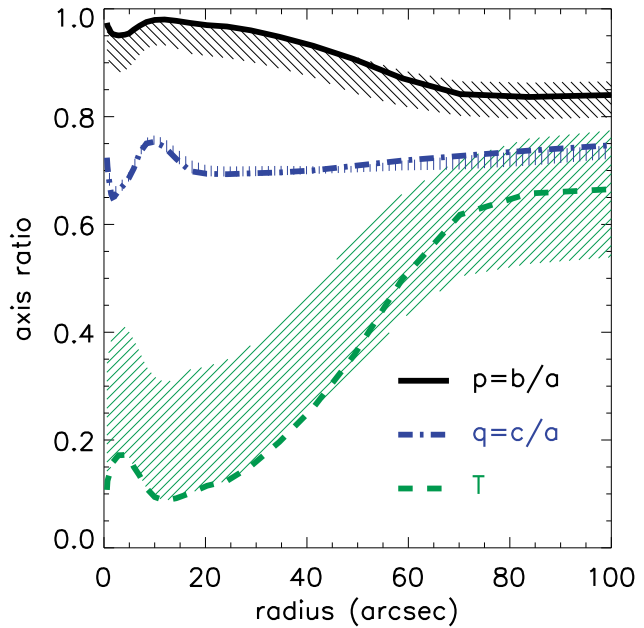}
\caption{Intrinsic axis ratios of the density distribution of the
  best-fit model of NGC~4365 as a function of radius. The inner
  35\arcs\ is nearly oblate axisymmetric ($p>0.95$), whereas the outer
  part is much more triaxial ($T>0.2$). The kinematic observations
  extend to $\sim35$\arcs.}
\label{figshape} 
\end{figure}

In Fig.~\ref{figshape}, we show the intrinsic shape parameters as a
function of radius of our best-fit model. Inside the central 35\arcs,
where we have the kinematic observations, the shape of NGC~4365 is
fairly oblate, with $p=\frac{b}{a}\ge0.95$, $0.65<q=\frac{c}{a}<0.75$
and the triaxiality parameter $T \equiv (1-p^2) / (1-q^2) <
0.2$. Further out -- outside of the area where observed kinematics are
available -- the reconstructed density becomes more prolate. This is
caused by the drop in $p$ to $\sim0.85$, while $q$ stays approximately
the same. As a result the triaxiality $T$ rises with increasing radius
to $\sim 0.6$.

The axis ratios found by \nocite{2004MNRAS.353....1S}{Statler} {et~al.} (2004) using velocity
field fitting of the \Sauron\ data are $\langle p\rangle\sim0.84$,
$\langle q\rangle\sim0.60$ which are just outside our 99\,\%
confidence, and do not agree with their claim of strong triaxiality
($\langle T\rangle\sim0.45$) inside 35\arcs\ of the galaxy. Their
lower limit on the triaxiality is not consistent with our measurement
in the center. We return to this below.

The corresponding best-fit viewing angles are
$(\vartheta,\varphi,\psi) = (68^\circ,73^\circ,91^\circ)$. To give an
indication of the uncertainty of these values we show a Lambert
azimuthal equal area projection of $\vartheta$ and $\varphi$ in
Fig.~\ref{pqugal}. This is comparable to a similar projection in
\nocite{2004MNRAS.353....1S}{Statler} {et~al.} (2004, their Fig.~4). Our best-fit viewing angle
is also not consistent with theirs. The velocity field fitting (VF)
method does not include the dispersion and other higher-order velocity
moments and assumes a plausible solution of the continuity equation,
while our Schwarzschild method does fit higher moments up to al least
$h_4$ and does not enforce any constraints on the DF.

The effect of the higher moments on the inferred shape can be studied
by making models without them. We find that excluding $h_4$ has no
significant effect on the recovered shape. Removing $h_3$ and/or lower
moments $h_2$ and $h_1$ (representing the dispersion and mean
velocity) changes the best-fit shape significantly and the
reconstructed LOSVD of these models show significant deviations from
Gaussian. These models are therefore not a good representation of the
observed LOSVD. This tests show that the Schwarzschild models need at
least $h_3$ to accurately recover the inferred shape and observed
LOSVD.

To compare our modeling to the VF method directly, we made models
where we only fit the mean velocity. We find that these models are
completely degenerate and no minimum could be found. Instead of our
least squares approach a likelihood method could be used to find a
solution in the probabilistic sense, however such methods are
unpractical for Schwarzschild models, because of the many iterations
required. The VF method does use Bayesian analysis, with a prior on
the dispersion.

For NGC~4365 to be a pure long-axis rotator with a KDC that consists
purely of short-axis tubes, we expect a best-fitting misalignment
angle $\psi$ that coincides with the observed kinematic misalignment
of $82^\circ \pm 2^\circ$ (or the symmetric 98\dgr). In fact, all the
models with $|\psi - 90| > 5$\dgr are strongly ruled out and we
conclude that NGC~4365 is not consistent with a pure long-axis
rotator.

\subsection{The orbital structure}

The observed kinematics that go out to $\sim35$\arcs\ show that
NGC~4365 must be intrinsically triaxial, due to the kinematically
decoupled core and the misaligned large scale rotation. This seems to
be in conflict with the (nearly) oblate axisymmetric shape inside
30\arcs of our best-fit model (Fig.~\ref{figshape}), as this shape
does not support the rotation around the major axis seen in the
observed kinematics.

\begin{figure}
\plotone{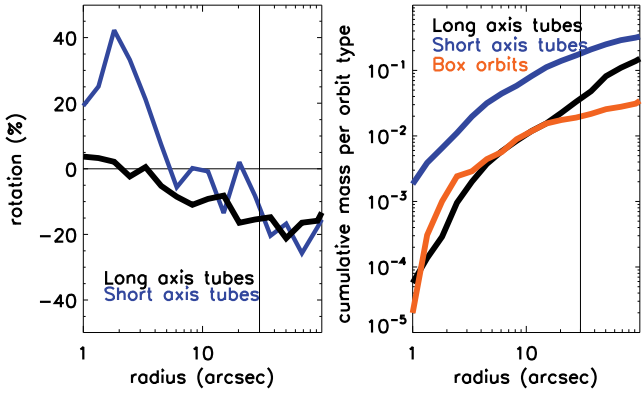}
\caption{\label{n4365orbtype}\label{n4365rot} Properties of the
orbital distribution. {\em Left:} Balance of prograde and retrograde
rotation as a function of radius. The balance is a fraction of the
total mass at that radius. The black vertical line represents the
radius beyond which we do not have kinematic observations from
\Sauron.  {\em Right:} Cumulative fraction of orbit type as function
of their start radius. }
\end{figure}

Fig.~\ref{n4365orbtype} shows the cumulative mass per orbit-type as a
function of intrinsic radius. As expected from the shape in the inner
region the stars on short axis tube orbits are dominant, accounting
for 75\% of the mass inside $30^{\prime\prime}$. The stars on long
axis tubes become significant in the model only outside
$30^{\prime\prime}$.

To understand the rotation seen in the observations we look at the
balance of stars on prograde and retrograde orbits, shown in
Fig.~\ref{n4365rot}. It shows that the stars in the KDC are on
prograde short axis orbits inside $6^{\prime\prime}$, while up to
$30^{\prime\prime}$ the stars on short axis orbits do not have a
preferred rotation direction. Only 15\% procent of the stars inside
$30^{\prime\prime}$ move on long axis orbits, but nearly all of them
move in the retrograde direction, and thus contribute to the observed
mean velocity.

\begin{figure}
\plotone{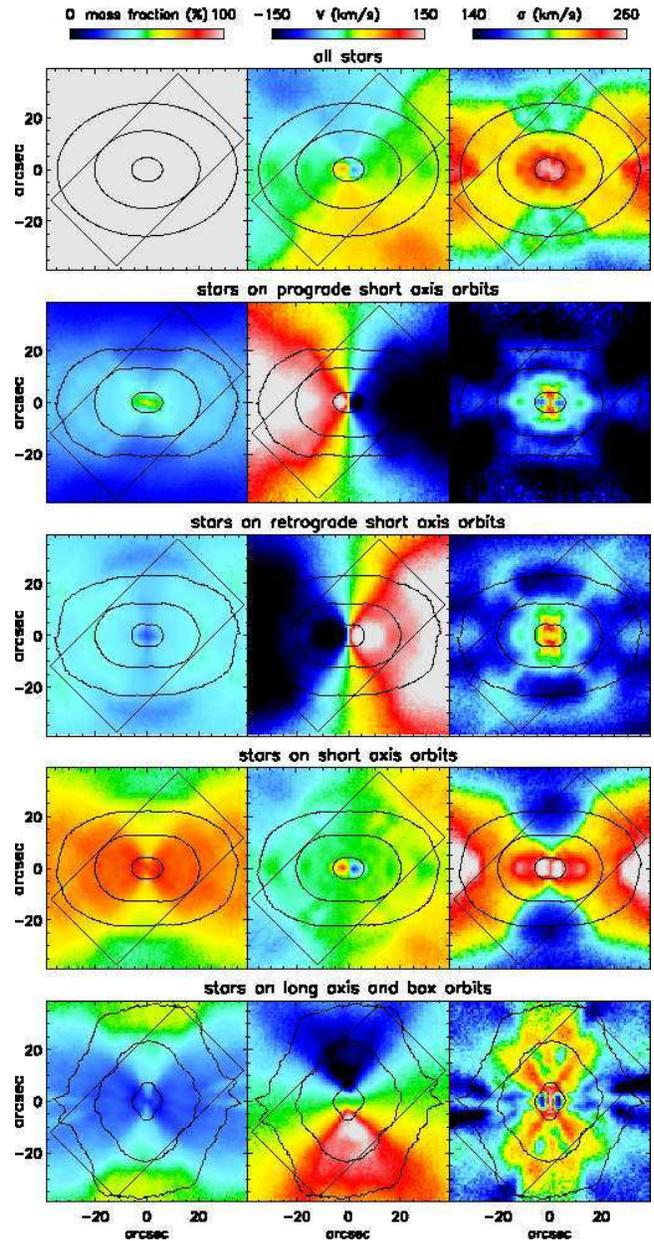}
\caption{\label{n4365muse} Extrapolated kinematics of the model of
  NGC~4365, showing (from top to bottom) the contribution from all
  stars, from stars on prograde, retrograde and combined short axis
  orbits, and from stars on long axis and box orbits. The contours
  show equal flux and the box shows the rectangle inside the \Sauron\ 
  observations.}
\end{figure}

To make the link between the orbital structure and the observations, we
show the unbinned kinematics of the stars on each type of orbit
individually, extrapolated over a region larger than the orginal
observations in Fig.~\ref{n4365muse}. The mass fraction, velocity
and dispersion field are shown for ($i$) all stars, ($ii$) stars on
prograde short axis orbits, ($iii$) stars on retrograde short axis orbits,
($iv$) stars on prograde and retrograde short axis orbits combined, and
($v$) stars on long axis and box orbits combined. 

In the decomposition of Fig.~\ref{n4365muse} the stars on prograde and
retrogrode short axis orbits have large velocities ($|v_{max}|>150$
\kms, $|\sigma|\sim160$ \kms) in opposite directions and the
combination of them lead to a velocity field with very little rotation
($|v_{max}|<60$ \kms) and a high dispersion field ($|\sigma|\sim220$
\kms). The stars on long axis orbits also rotate quickly
($|v_{max}|>150$ \kms) and although they contribute only $\sim20$\% of
the mass, the large velocity adds significantly to the large scale
rotation seen in the observations.

Two other galaxies NGC~4550 and NGC~4473 show very similar features to
NGC~4365, both have unusually high sigma along the major axis
\nocite{2004MNRAS.352..721E}({Emsellem} {et~al.} 2004).  They both consist of two counterrotating
disks, similar to the two main components in NGC~4365: the prograde
and retrograde short axis orbits \nocite{1992ApJ...400L...5R, PaperX}({Rix} {et~al.} 1992; {Cappellari} {et~al.} 2007).

\begin{figure*}
\plottwo{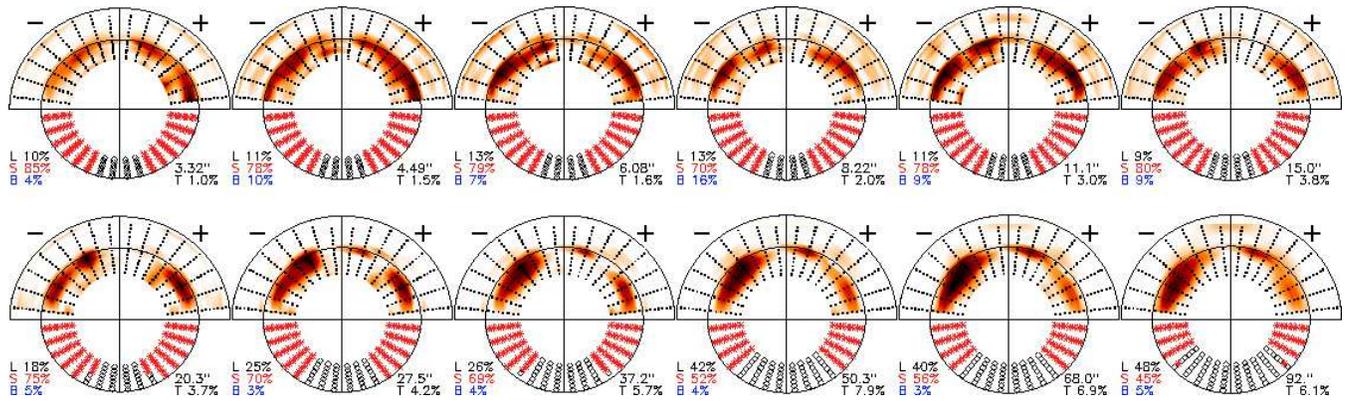}
\caption{\label{n4365df} The orbits that receive weight in the
  best-fitting model of NGC~4365. The top part of each panel is the
  same as in Fig.~\ref{weights} and the bottom half is as in
  Fig.~\ref{xz_startspace}.  L,S,B and T denote Long axis tubes, Short
  axis tubes, Boxes and percentage of Total mass, respectively. For
  each panel, the corresponding radius is given at the lower right, in
  units of arcsec.}
\end{figure*}

The KDC is not very distinct in the decomposition
(Fig.~\ref{n4365muse}) and is perhaps only an appearance. The only way
to really disentangle the KDC is through the unbalance of the central
stars on prograde short axis orbits (Fig.~\ref{n4365rot}). However, in
the orbital weights (Fig.~\ref{n4365df}) the KDC seems an integral
part of all the stars on prograde short axis tube orbits as these stars are
continuously distributed in orbital weights. It is therefore difficult
to see the KDC as a distinct kinematic component.

\section{Discussion and Conclusion}
\label{discussion}

We have presented a flexible method to build dynamical models of
triaxial early-type galaxies, allowing for position angle twist and
ellipticity variation in their surface brightness, as well as a
central supermassive black hole. The method is based on
Schwarzschild's orbit superposition technique and uses the observed
surface brightness distribution and observed kinematics to make a
triaxial model of the observed galaxy.  Our models can be constrained
by observations of the full line-of-sight velocity distribution.

We discussed tests of our method on the triaxial Abel model with a
separable potential from vdV07, which illustrates the accuracy of our
orbit classification method and the effect of regularisation. Tests
with the viewing angles showed that we can constrain the intrinsic
shape of galaxies with significant structure in their velocity field.

We also presented results of an application on two-dimensional data of
the E3 galaxy NGC~4365, obtained with the integral-field spectrograph
\Sauron. We showed that our method is capable of reproducing the main
observational features to within the errors. We found a best-fitting
($I$-band) stellar mass-to-light ratio of $3.5\pm0.2$ in solar units
and best-fitting viewing angles of ($\vartheta,\varphi,\psi$)=
($68^\circ,73^\circ,91^\circ$). The characteristic axis ratios are
$p\ge0.95$, $0.65<q<0.75$ inside $35^{\prime\prime}$. By applying a
simple regularisation scheme, we were able to determine the
distribution of orbital weights, which provided us with a view on the
orbit structure of this galaxy. We find the inner part to be nearly
oblate axisymmetric, with most of the stars equally divided on
prograde and retrograde short axis tube orbits. 
Further out the galaxy becomes more triaxial, and the stars orbit on
both long axis and short axis tubes. The KDC seen in the observations
is not dynamically distinct from the main body of the galaxy. More
evidence for the idea that the `decoupled core' is part of the main
body of the galaxy, comes from the stellar ages determined by
\nocite{2001ApJ...548L..33D}{Davies} {et~al.} (2001). The ages of the stars where determined to
be at least $\sim$12 Gyr and they do not show a strong dependence
which radius. Overall our orbital structure is consistent with the
results from \nocite{1991AJ....102..882S}{Statler} (1991) and
\nocite{1994MNRAS.271..924A}{Arnold} {et~al.} (1994).

An important consideration is the stability of triaxial galaxies. A
significant fraction of the centrophilic box orbits can become chaotic
in the presence of a central cusp or a supermassive black hole
\nocite{1985MNRAS.216..467G, 1998ApJ...506..686V}({Gerhard} \& {Binney} 1985; {Valluri} \& {Merritt} 1998). As box orbits are
crucial for supporting the triaxial shape, it is not evident whether a
triaxial object with a central black hole can retain its shape over a
Hubble time \nocite{1983ApJ...270...51L}({Lake} \& {Norman} 1983). Earlier $N$-body simulations
of triaxial galaxies in which a central mass concentration is grown
indeed show a fairly rapid evolution towards a rounder shape in the
inner parts \nocite{1998ApJ...498..625M, 1998ApJ...506..686V}(e.g. {Merritt} \& {Quinlan} 1998; {Valluri} \& {Merritt} 1998),
but these results were challenged recently \nocite{2002ApJ...567..817H,
  2002ApJ...568L..89P}({Holley-Bockelmann} {et~al.} 2002; {Poon} \& {Merritt} 2002). The intrinsic shape of our best-fit model is
nearly oblate axisymmetric in the centre, and more triaxial further
out. This might be a sign of evolution towards a axisymmetric shape,
induced by the central massive black hole from the inside out.
Clearly, we need to obtain a better understanding of whether triaxial
galaxies can reach stationary equilibrium and if not, what the
time-scale of the transition toward a nearly spheroidal shape is.

The extension from an axisymmetric to a triaxial implementation of
Schwarzschild's method opens up a wide range of applications. For
example, while the mass of the central black hole seems to be
correlated with other properties of the galaxy
\nocite{2000ApJ...539L...9F, 2000ApJ...539L..13G, 2002ApJ...574..740T}({Ferrarese} \& {Merritt} 2000; {Gebhardt} {et~al.} 2000; {Tremaine} {et~al.} 2002),
nearly all black hole estimates were derived with dynamical models of
(edge-on) axisymmetric models. In such models, the increase in
line-of-sight motion towards the center that is seen in
high-resolution observations of many nearby galaxies can indeed only
be explained by a black hole. In triaxial systems, box orbits provide
an alternative way of creating motion along the line-of-sight
(depending on the viewing angles, see e.g.,
\nocite{1988MNRAS.232P..13G}{Gerhard} 1988). As a result, a given observed velocity
profile may require a different black hole mass when the galaxy is
allowed to be triaxial, which in turn may influence the black hole
mass correlations if the intrinsic shape correlates with luminosity,
as has been suggested.

Another area of interest is the validation of predictions on the halo
shapes from galaxy merger \nocite{2002ApJ...568...52W}(e.g., {Wechsler} {et~al.} 2002) and
$\Lambda$CDM cosmology simulations
\nocite{2000ApJ...538..477N}(e.g., {Navarro} \& {Steinmetz} 2000). They predict the existence of
strongly triaxial halos as a result of merging. To confirm these
simulations, our method can be used to measure the halo shapes of a
representative sample of galaxies, using kinematical observation at
large radii, where the halo mass dominates.

Our method can be further extended in a number of aspects. For
example, we assumed that the galaxy as a whole is non-rotating. The
reason for this is that inclusion of figure rotation further
complicates matters \nocite{1982ApJ...263..599S, 1982ApJ...258..490H}({Schwarzschild} 1982; {Heisler}, {Merritt} \&  {Schwarzschild} 1982),
while it may not be crucial for the modeling of existing observations
of giant elliptical galaxies. The fitting of kinematics can be
improved by fitting the LOSVD directly or by using eigen-velocity
profiles \nocite{2006MNRAS.367....2H}({Houghton} {et~al.} 2006). Additionally, the method can be
enhanced to include line strength information and multiple stellar
populations, to study the distribution of stellar ages and
metallicities within galaxies. Even without additional extensions, the
triaxial Schwarzschild method allows us to investigate the intrinsic
structure and orbital make-up of early-type galaxies.

\section*{Acknowledgements}

It is a pleasure to thank Eric Emsellem, Karl Gebhardt and Roeland van
der Marel for a critical reading of an earlier version of the
manuscript. We also like to thank Jes\'us Falc\'on-Barroso for the
image of NGC~4365.

The \Sauron\ project is made possible through grants 614.13.003,
781.74.203, 614.000.301 and 614.031.015 from NWO and financial
contributions from the Institut National des Sciences de l'Univers,
the Universit\'e Claude Bernard Lyon~I, the Universities of Durham,
Leiden, and Oxford, the British Council, PPARC grant `Extragalactic
Astronomy \& Cosmology at Durham 1998--2002', and the Netherlands
Research School for Astronomy NOVA. GvdV acknowledges support provided
by NASA through grant NNG04GL47G and through Hubble Fellowship grant
HST-HF-01202.01-A awarded by the Space Telescope Science Institute,
which is operated by the Association of Universities for Research in
Astronomy, Inc., for NASA, under contract NAS 5-26555. MC acknowledges
support from a VENI grant 639.041.203 awarded by the Netherlands
Organization for Scientific Research (NWO). Part of this work is based
on data obtained from the ESO/ST-ECF Science Archive Facility.


\appendix

\label{lastpage}
\end{document}